\documentclass[a4paper,11pt]{article}

\usepackage{jheppub}
\usepackage{mathrsfs}

\renewcommand{\d}{\textrm{d}} 
\renewcommand{\L}{\mathscr{L}} 
\newcommand{\N}{\mathcal{N}} 
\newcommand{\D}{\mathcal{D}} 
\newcommand{\covD}{\mathscr{D}} 
\newcommand{\W}{\mathcal{W}} 
\renewcommand{\O}{\mathcal{O}} 
\newcommand{\alphadot}{{\dot\alpha}}
\newcommand{\hc}{\textrm{h.c.}}
\DeclareMathOperator{\tr}{tr}
\newcommand{\eomequal}{\stackrel{\text{e.o.m.}}{=}}

\title{Spontaneous conformal symmetry breaking \\
and a massless Wu-Yang monopole}

\author{Marc Gillioz}

\affiliation{Theoretical Particle Physics Laboratory, Institute of Physics, EPFL, Lausanne, Switzerland}

\emailAdd{marc.gillioz@epfl.ch}

\date{\today}

\abstract{
A formulation of $\N = 2$ supersymmetric Yang-Mills theory with a spacetime-dependent gauge coupling allows to study the breaking of conformal symmetry at the quantum level. The theory has an energy-momentum tensor that is only conserved if an equation of motion for the coupling is imposed. It admits non-trivial solitons, among which the Wu-Yang monopole that can be regularized and turns out to be massless. On the other hand, the ordinary BPS monopole is only a solution in the large $N_c$ limit.
}

\begin{document} 
\maketitle
\flushbottom


\section{Introduction}
\label{sec:introduction}

Most quantum field theories are not invariant under scale transformations, even when their Lagrangian does not depend on any dimensionful parameter:
this is a well-understood consequence of the presence of ultraviolet (UV) divergences in the quantum theory, which need to be regularized.
One of the most compelling approaches to quantum field theory is the Wilsonian renormalization group (RG), in which the theory is endowed with a physical cutoff suppressing high-frequency modes and leading to UV-finite physical observables. The theory can be made invariant under a change in the cutoff scale provided that the couplings acquire an implicit dependence on the cutoff itself.
In this way, the quantum effective action possesses an exact scale symmetry that is only broken once the value of the couplings is fixed in a specific process.
The Callan-Symanzik equation is simply the Ward identity associated to this symmetry.
The Wilsonian quantum effective action realizes therefore the mechanism of spontaneous breaking of scale invariance.%
\footnote{In practice, one often uses a different approach to renormalization, in which the couplings do not depend on a cutoff, but rather on a physical quantity of the process under consideration, such as the momentum exchange in a scattering experiment.
An unpleasant feature of this alternative approach is that it necessarily requires using an effective action constructed from a non-local Lagrangian.}

Unitary quantum field theories that possess a scale symmetry are known to be also invariant under the larger symmetry group of conformal transformations, which is generated by a \emph{local} rescaling of distances.%
\footnote{To be precise, arbitrary local rescalings of distances generate Weyl transformations, of which conformal transformations are only a subset; in any case, the two concepts are intimately linked for unitary theories, see for instance Refs.~\cite{Karananas:2015ioa, Farnsworth:2017tbz}.}
Even though the link between these two symmetries is not completely understood \cite{Polchinski:1987dy, Luty:2012ww, Dymarsky:2014zja}, it seems to arise naturally from the point of view of the Wilsonian renormalization group: \emph{a priori}, there is no obstruction to using a different physical cutoff scale at each point in space and/or time, provided that the fluctuations are smooth enough. The dependence of the couplings on the cutoff required for physical observables to be independent of the latter means that all the coupling constants of the theory must be promoted to fields, an idea that is at the core of the so-called \emph{local renormalization group}~\cite{Drummond:1977dg, Shore:1986hk, Tseytlin:1986ws, Osborn:1987au}.
The historical motivation for the development of the local RG is that it allows for a simple definition of composite operators, without the need for additional local counterterms, since the coupling fields act as sources for them. The great success of the local RG is a derivation of the equivalent of Zamolodchikov's \emph{C}-theorem in four spacetime dimensions~\cite{Osborn:1989td, Jack:1990eb, Osborn:1991gm, Jack:2013sha, Shore:2016xor, Baume:2014rla}. More generally, the use of the local renormalization group allows to consider a quantum effective action that possesses the full conformal symmetry, only (spontaneously) broken by a choice of physical values for the coupling fields.

In the literature, the local RG is merely used as a technical tool and the couplings are set to constant values at the end of the analysis.
The reason for this is very simple: a cutoff that is not constant breaks explicitly the translation symmetry, a feature that is usually not welcome in a quantum field theory.
Among other obstacles, a non-constant cutoff --- or equivalently non-constant renormalized couplings --- implies that the theory does not have a conserved energy-momentum tensor: the Noether current $T^{\mu\nu}$ associated to translations obeys
\begin{equation}
	\partial_\nu T^{\mu\nu} \propto \partial^\mu g \,
	\frac{\delta \Gamma}{\delta g},
	\label{eq:T:teaser}
\end{equation}
where $\Gamma$ is the quantum effective action and $g$ a coupling field. A constant cutoff also seems natural from a purely mathematical perspective: since conformal symmetry is broken spontaneously, it should occur in a way that preserves the largest possible subgroup of symmetries.
However, this subgroup does not necessarily have to be the Poincar\'e group of spacetime isometries, but could be any other 10-dimensional subgroup of the 16-dimensional conformal group in four dimensions~\cite{Fubini:1976jm}.
Moreover, there are physical systems that break translation invariance explicitly, such as an isolated soliton.

The goal of this paper is precisely to study the existence of solitons in a quantum field theory with a spacetime-dependent cutoff. For simplicity, this work focuses exclusively on $\N = 2$ supersymmetric pure gauge theory, which has all the characteristics of an interesting theory --- a strongly-coupled regime and confinement in the IR --- and yet is simple enough that the quantum effective action does not get perturbative corrections beyond one loop.
The local renormalization group approach is also particularly powerful when combined with supersymmetry~\cite{Kaplunovsky:1994fg, Freedman:1998rd, Osborn:2003vk, Auzzi:2015yia}. Ignoring the $\theta$-angle of the vacuum, the supersymmetric $SU(N_c)$ gauge theory is characterized by a single coupling which we denote by $g$, and consistency under the local renormalization group requires to enhance the ordinary quantum effective action with local terms built out of derivatives of $g$, namely~\cite{Gillioz:2016ynj}
\begin{equation}
	\L \supset \frac{N_c^2-1}{16 \pi^2} \,
	\left[ \frac{1}{2} \, \left( \square g \right)^2
	+ \frac{1}{6} \left( \partial_\mu g \, \partial^\mu g \right)^2 \right].
	\label{eq:L:teaser}
\end{equation}
The conformal invariance of this effective action implies an inherent arbitrariness in the value of the coupling field $g$. In a semi-classical approach, however, one is interested in solitonic field configurations that are stationary points of the action, i.e.~for which the classical equations of motions for the fields are satisfied. This involves imposing an equation of motion for the coupling field $g$ as well as for other fields, 
\begin{equation}
	\frac{\delta \Gamma}{\delta g} = 0.
	\label{eq:eom:teaser}
\end{equation}
In the absence of terms such as eq.~\eqref{eq:L:teaser}, this condition would simply account for setting to zero all interaction terms in the Lagrangian; instead, eq.~\eqref{eq:eom:teaser} turns out to be a non-trivial equation that can potentially lead to stable soliton solutions. It also provides the condition necessary to make sense of a soliton's energy, as the energy-momentum tensor of eq.~\eqref{eq:T:teaser} is precisely conserved upon eq.~\eqref{eq:eom:teaser}.

The results presented in this paper are non-exhaustive: they are meant to illustrate the power of the method that consists in using spacetime-dependent couplings in the search of solitons. Nevertheless, two important observations emerge. The first one is that the ordinary monopole and dyon solutions  obtained at constant coupling~\cite{'tHooft:1974qc, Polyakov:1974ek, Prasad:1975kr, Bogomolny:1975de, Julia:1975ff} are \emph{not} solutions to the equation of motion~\eqref{eq:eom:teaser} in general, but only in the large $N_c$ limit; a more general solution at finite $N_c$ could be obtained in a perturbative expansion in inverse powers of $N_c$, but the procedure is delicate and not pursued in the present work. The other important result is the observation that the ancient Wu-Yang monopole~\cite{Wu:1975vq} is on the contrary a solution to eq.~\eqref{eq:eom:teaser} at arbitrary $N_c$; and while it was historically discarded as a physical solution due to a divergence in its energy density, its semi-classical mass can be regularized and vanishes precisely when using the approach involving spacetime-dependent couplings.
The physical interpretation of this massless monopole is potentially very interesting, as S-duality suggests the existence of such particles in the unbroken phase of the gauge theory~\cite{Seiberg:1994rs, Lee:1996vz}. A hypothetical semi-classical description of monopole condensation certainly requires a complete study of all solitons and instantons of the theory, which goes beyond the scope of this work.
Nevertheless, the massless Wu-Yang monopole is interesting in its own right.

Finally, let us recall that we have referred so far to the spacetime-dependent cutoff as an input to the derivation of the quantum effective action; it could be tempting to interpret the coupling $g$ as a dynamical field, in which case the equation of motion~\eqref{eq:eom:teaser} follows from the path integral formulation. The field $g$ could then be interpreted as a dilaton, the Goldstone boson of spontaneous conformal symmetry breaking~\cite{Nielsen:1975hm, Low:2001bw}, and eq.~\eqref{eq:L:teaser} seen as the dilaton effective action. Even though this interpretation is satisfactory in many aspects, it is deeply pathological due to the presence of ghosts~\cite{Nicolis:2008in, Nicolis:2009qm}.

The paper is organized as follows.
We begin in Section~\ref{sec:effectiveaction} by recalling the results of Ref.~\citep{Gillioz:2016ynj} about the quantum effective action of $\N = 2$ Yang-Mills theory with spacetime-dependent couplings.
In Section~\ref{sec:conformalsymmetry}, we examine the anomaly that appears in the theory due to these non-constant couplings and show that it vanishes when imposing an equation of motion for them, hence realizing conformal symmetry at the quantum level. 
Section~\ref{sec:semiclassics}, is devoted to the semi-classical analysis of the theory, including a discussion of the boundedness of the energy, the derivation of the massless Wu-Yang monopole solution and of the ordinary BPS monopole solution at large $N_c$.
The details of the computation of the equations of motion and of the energy-momentum tensor are left for Appendix~\ref{sec:supercurrent}.


\section{The quantum effective action}
\label{sec:effectiveaction}

This section is a short review of Ref.~\cite{Gillioz:2016ynj}, a derivation of the quantum effective action of $\N = 2$ supersymmetric Yang-Mills theory without matter fields. It makes use of the $\N = 1$ superspace language, with standard notation given for completeness in Appendix~\ref{sec:supercurrent}.%
\footnote{The use of $\N = 2$ superfield methods could seem more appropriate, but the $\N = 1$ formulation contains all the ingredients needed in this work, namely supergraph methods to compute the effective action and non-renormalization theorems following from holomorphy.}

The Wilsonian quantum effective action is precisely the right object to study in a search for solitons, as it is local, has an obvious semi-classical limit, and does not depend on the infrared subtleties that affect for instance the one-particle irreducible effective action.
The starting point is the observation of Ref.~\cite{ArkaniHamed:1997mj} that there exist a holomorphic regularization scheme for supersymmetric pure gauge theories, in which the theory is defined from the $\N = 4$ supersymmetric action deformed with a soft mass term for the chiral matter multiplets. The effective theory obtained when integrating out the massive superfields with mass $\Lambda$ is precisely the Wilsonian quantum effective action for the $\N = 1$ theory, with an explicit dependence on $\Lambda$, which plays the role of the cutoff scale.
Since the $\N = 4$ theory is finite and the breaking is soft, only finite supergraphs enter the computation of the effective action.

The $\N = 2$ pure gauge action can be obtained in a similar fashion by using a mass deformation for two of the three chiral multiplets of the $\N = 4$ theory only: the remaining massless fields are the gauge multiplet $V$ and one chiral multiplet $\Phi$, forming together an $\N = 2$ extended multiplet. A one-loop computation shows that the dependence of the cutoff $\Lambda$ is logarithmic and can absorbed into a redefinition of the holomorphic gauge coupling $\tau_0$ of the $\N = 4$ theory into $\tau[\Lambda]$, defined by
\begin{equation}
	\tau_0 \to \tau[\Lambda] =  \tau_0
	+ \frac{3 N_c}{8\pi^2} \log\left( \frac{\Lambda}{M} \right)
	\equiv \frac{3 N_c}{8\pi^2} \log\left( \frac{\Lambda}{\tilde{M}} \right),
	\label{eq:tau:cutoff}
\end{equation}
where the gauge group is $SU(N_c)$ and $M$ is an arbitrary mass scale. There are no higher-order perturbative corrections, and the action of the $\N = 2$ gauge theory with $\tau[\Lambda]$ as the holomorphic coupling is therefore exact in perturbation theory.
The holomorphic beta function can be read directly from eq.~\eqref{eq:tau:cutoff}: it is one-loop exact and constant, given by the loop coefficient $b = 3N_c/8\pi^2$.

The idea developed in Ref.~\cite{Gillioz:2016ynj} is to promote the mass-deformation $\Lambda$ to a chiral superfield and to repeat the derivation of the quantum effective action. In addition to the logarithmic dependence in $\Lambda$ obtained in Eq.~\eqref{eq:tau:cutoff}, the effective action can now depend on supersymmetric covariant derivatives $\D_\alpha$ and $\bar{D}_\alphadot$ of the chiral superfield $\Lambda$ and its hermitian conjugate $\bar{\Lambda}$. The cutoff scale of the effective action is set by the vacuum expectation value of $\Lambda$, which we can take to be very large. It is natural to assume that the fluctuations of $\Lambda$ around its central values are small, and the effective action can therefore be obtained in a derivative expansion. After a careful derivation, the only relevant terms in this expansion are D-terms with four supersymmetric covariant derivatives of $\Lambda$. There are infinitely many of them, but they can be resummed \emph{\`a la} Coleman-Weinberg, and the quantum effective action of the $\N = 2$ theory is finally given by the Lagrangian
\begin{eqnarray}
	\L & = & \frac{b}{8} \int \d^2\theta \,
	\log\left( \frac{\Lambda}{\tilde{M}} \right) \tr(\W^\alpha \W_\alpha)
	+ b \int \d^4\theta \log\left( \frac{\Lambda}{\tilde{M}} \right)
	\tr \left( \bar{\Phi} e^{-2V} \Phi e^{2V} \right)
	+ \hc
	\nonumber \\
	&& + \frac{c}{2} \int \d^4\theta \,
	\frac{\partial_\mu \Lambda \, \partial^\mu \bar{\Lambda}}
	{\Lambda \bar{\Lambda}}
	+ \frac{c}{48} \int \d^4\theta \,
	\frac{\D^\alpha \, \Lambda \D_\alpha \Lambda \,
	\bar{\D}_\alphadot \bar{\Lambda} \, \bar{\D}^\alphadot \bar{\Lambda} }
	{\Lambda^2 \bar{\Lambda}^2}.
	\label{eq:L:superspace:cutoff}
\end{eqnarray}
where $\D$ and $\bar{\D}$ are the supersymmetric covariant derivatives defined in eq.~\eqref{eq:covariantderivatives}, $\W_\alpha$ is the holomorphic field strength tensor~\eqref{eq:W}, and $b$ and $c$ are real, positive group-theoretical coefficients corresponding to
\begin{equation}
	b = \frac{3 N_c}{8\pi^2},
	\hspace{1.5cm}
	c = \frac{N_c^2 - 1}{16 \pi^2}.
\end{equation}
Interestingly, this action is local and does not depend explicitly on the arbitrary renormalization scale $\tilde{M}$, in the sense that we can define a dimensionless chiral superfield
\begin{equation}
	G = \log\left( \frac{\Lambda}{\tilde{M}} \right),
	\label{eq:G:def}
\end{equation}
for which the action~\eqref{eq:L:superspace:cutoff} becomes
\begin{eqnarray}
	\L & = & \frac{b}{8} \int \d^2\theta \,
	G \, \tr\big( \W^\alpha \W_\alpha \big)
	+ b \int \d^4\theta \, G \,
	\tr \big( \bar{\Phi} e^{-2V} \Phi e^{2V} \big)
	+ \hc
	\nonumber \\
	&& + \frac{c}{2} \int \d^4\theta \,
	\partial_\mu G \, \partial^\mu \bar{G}
	+ \frac{c}{48} \int \d^4\theta \,
	\D^\alpha G \, \D_\alpha G \, \bar{\D}_\alphadot \bar{G} \,
	\bar{\D}^\alphadot \bar{G}
	\label{eq:L:superspace}
\end{eqnarray}
Note that this action is not actually invariant under extended $\N = 2$ supersymmetry: the use of a $\N = 1$ chiral superfield as the cutoff spoils half of the supersymmetries. A complete $\N = 2$ invariant formulation is possible, but not needed in any aspect of this work. To be more precise, in the following sections we will work mostly with the action in components, and only make use of the real part $g$ of the lowest component of $G$ and $\bar{G}$. The extended supermultiplet that would play the role of the cutoff in the $\N = 2$ formulation consists in two $\N = 1$ chiral multiplets, of which the lowest components can be arranged in a R-symmetry singlet and a triplet. The field $g$ is the R-symmetry singlet, and thus interacts only with $\N = 2$ supersymmetry-preserving terms in the action. Explicitly, the action in components reads
\begin{align}
	\L = 
	- \frac{b \, g}{2} \, & \left[
	\tr\big( F_{\mu\nu} F^{\mu\nu} \big)
	+ \tr\big( \phi^\dag \, D_\mu D^\mu \phi \big)
	+ \tr\big( \phi \, D_\mu D^\mu \phi^\dag \big)
	+ \frac{1}{2} \, \tr\big( [ \phi^\dag, \phi ]^2 \big) 
	+ \ldots \right]
	\nonumber \\
	+ c \, &
	\left[ \frac{1}{2} \, \left( \square g \right)^2
	+ \frac{1}{6} \, \left( \partial_\mu g \, \partial^\mu g \right)^2 \right]
	+ \ldots,
	\label{eq:L}
\end{align}
where $F_{\mu\nu}$ and $D_\mu$ are respectively the field strength tensor and the covariant derivative associated with the $SU(N_c)$ gauge group, and $\phi$ is a complex scalar transforming in the adjoint representation of $SU(N_c)$.
Note that $g$ does not correspond the ordinary gauge coupling, but rather to the real part of the holomorphic coupling normalized with an unusual loop factor.
The ellipses in the first line of eq.~\eqref{eq:L} indicate the fermionic fields that do not play any role in our analysis, and in the second line the superpartners of the coupling field $g$ that we choose to set to zero. Appendix~\ref{sec:supercurrent} contains an complete expression for the Lagrangian in components, as well as an argument explaining why the superpartners of $g$ can be ignored.
Note that the form of the kinetic term for $\phi$ is not arbitrary: it is completely fixed once the freedom of integration by parts has been used to get rid of derivatives acting on $g$.

The Lagrangian~\eqref{eq:L} is the starting point of our analysis, and the reader who does not care about supersymmetry and holomorphic renormalization schemes can take it as granted and work his/her way through the rest of the paper without difficulties.
Notice however that the $\N = 1$ superspace formulation~\eqref{eq:L:superspace} was instrumental in the derivation of the terms involving derivatives of $g$, and that it will still prove useful in determining the energy-momentum tensor of the theory.
Let us finally remark that the terms proportional to $c$ are not completely new, as they are the terms needed to reproduce the scale anomaly obtained in Ref.~\cite{Jack:1990eb} for gauge theories, even though the exact correspondence is subtle and will not be discussed further here. We can at most comment that the sign of these terms is consistent with the existence of an \emph{a}-theorem establishing the monotonicity of the renormalization group flow~\cite{Komargodski:2011vj, Komargodski:2011xv}.%
\footnote{More precisely, the coefficient of the $(\square g)^2$ term seems to be related to the flow of the $\square R$ anomaly in curved space, for which there also exist some monotonicity properties~\cite{Anselmi:1999xk, Prochazka:2017pfa}.}
For us, this sign will simply mean that a non-constant gauge coupling $g$ gives a negative contribution to the mass of a soliton, leading to stable solutions.

We move on to study the properties of the quantum effective action of $\N = 2$ supersymmetric Yang-Mills with spacetime-dependent coupling.
Before discussing the presence of new solitons in this theory, we elaborate on the important fact that the action presented in this section contains all the information on the renormalization group even at the classical level.


\section{Conformal symmetry and its breaking}
\label{sec:conformalsymmetry}

In a traditional treatment of quantum field theory with constant couplings, the definition of composite operators does not follow straightforwardly from the quantum effective action: when using auxiliary source field for composite operators in the generating functional, one should be aware of the need of adding additional counterterms with contact interactions for the sources, so as to preserve the finiteness of correlation functions of these composite operators. These counterterms play a key role in the breaking of scale (and conformal) symmetry in the quantum theory, but they are not visible at first sight. This is the reason why for instance the quantum effective action of Yang-Mills theory appears to be conformally invariant, even though it is well known that the theory is not, due to the presence of the trace anomaly.

One of the main virtues of the quantum effective action with spacetime-dependent couplings is to avoid the requirement of these additional counterterms, and therefore to make the breaking of conformal invariance explicit: the physics of composite operators is essentially included in the action since the coupling fields act as sources for them. To illustrate this, let us consider a scale transformation on the action~\eqref{eq:L}, acting as a simultaneous rescaling of the metric tensor $\eta_{\mu\nu}$ and of the fields
\begin{equation}
	\delta_\sigma \eta_{\mu\nu} = 2 \, \sigma \, \eta_{\mu\nu},
	\hspace{1.5cm}
	\delta_\sigma \O(x) = d_\O \, \sigma \, \O(x),
	\label{eq:Weylscaling}
\end{equation}
where $\O$ denotes any field of the theory and $d_\O$ its scaling dimension. A notable exception to this transformation rule for the fields applies to the coupling $g(x)$, which transforms instead linearly,
\begin{equation}
	\delta_\sigma g(x) = \sigma,
	\label{eq:Weylscaling:g}
\end{equation} 
following the definition~\eqref{eq:G:def} in terms of the cutoff field that has scaling dimension of a mass term: $\delta_\sigma \Lambda = \sigma \, \Lambda$. In a theory with only marginal operators, such as Yang-Mills, this implies that the overall variation of the quantum effective action follows
\begin{equation}
	\delta_\sigma \Gamma = \frac{\delta \Gamma}{\delta g},
	\label{eq:Weylvariation}
\end{equation}
which is non-zero in general, even when the coupling are set to constant values \emph{a posteriori}. Since the variation on the left-hand side is related to the trace of the energy-momentum tensor and the functional derivative on the right-hand side is the definition of a composite operator, the above statement is precisely equivalent to the usual relation
\begin{equation}
	[T^\mu_{~\mu}(x)] = \sum_\O \beta_\O \, [\O(x)]
	\label{eq:traceanomaly}
\end{equation}
where $\beta_\O$ is the beta function for the coupling of the marginal operator $\O$, and the square brackets indicates that this equation is valid for renormalized operators. These remarks apply to arbitrary quantum field theories, but let us now focus specifically on $\N = 2$ supersymmetric Yang-Mills and illustrate the crucial advantage of using spacetime-dependent couplings.

\subsection{An anomaly in superspace}

In a supersymmetric theory, classical conformal invariance of the effective action means that there exists a superfield $J_{\alpha\alphadot}$ that is conserved classically, i.e.~a current that vanishes upon the superspace equations of motion when taking its divergence with supersymmetric covariant derivatives,
\begin{equation}
	\D^\alpha J_{\alpha\alphadot} \eomequal 0,
	\hspace{1.5cm}
	\bar{\D}^\alphadot J_{\alpha\alphadot} \eomequal 0.
	\label{eq:supercurrent:conservation}
\end{equation}
The lowest component of this supercurrent is the $R$-current, but most importantly its $\theta\bar{\theta}$ component corresponds to the classical energy-momentum tensor $T^{\mu\nu}$, automatically conserved by the above constraint.
Considering the simplest example of a pure $\N = 1$ non-abelian gauge theory, the supercurrent is of the form
\begin{equation}
	J_{\alpha\alphadot} = \tr\big( \W_\alpha e^{2V} \bar{\W}_\alphadot e^{-2V} \big).
\end{equation}
It is the unique gauge-invariant operator whose lowest component has the scaling dimension of a conserved current, up to an overall normalization constant. Conservation of this current follows indeed from the classical equation of motion
\begin{equation}
	\covD^\alpha \W_\alpha \eomequal 0,
	\qquad \Leftrightarrow \qquad
	\bar{\covD}_\alphadot \bar{\W}^\alphadot \eomequal 0,
\end{equation}
where $\covD$ indicates the gauge-covariant derivative.
The non-vanishing trace of the energy-momentum tensor is hidden in this formalism, as it only arises from the fact that the classical supercurrent must be distinguished from its renormalized counterpart.

If one consider instead the same $\N = 1$ gauge theory but replace the holomorphic coupling with a superfield $G$, the superspace equation of motion gets modified to
\begin{equation}
	\big( G + \bar{G} \big) \, \covD^\alpha \W_\alpha
	+ \big( \D^\alpha G \big) \, \W_\alpha
	+ \big( \bar{\D}_\alphadot \bar{G} \big) \, 
	e^{2V} \bar{\W}^\alphadot e^{-2V} 
	\eomequal 0.
\end{equation}
The candidate supercurrent would then be of the form
\begin{equation}
	J_{\alpha\alphadot} = \left( G + \bar{G} \right) \,
	\tr\big( \W_\alpha e^{2V} \bar{\W}_\alphadot e^{-2V} \big),
\end{equation}
but it is not conserved, 
\begin{equation}
	\D^\alpha J_{\alpha\alphadot} \eomequal 
	-\frac{1}{2} \,
	\bar{\D}_\alphadot \bar{G} \,
	\tr\big(\bar{\W}_{\dot\beta} \bar{\W}^{\dot\beta} \big),
	\label{eq:J:nonconservation}
\end{equation}
unless the coupling field is constant.
No classically conserved supercurrent can be written down for this simple theory, nor for the $\N = 2$ theory with an additional chiral multiplet $\Phi$. This observation was first made in Refs.~\cite{Kraus:2001tg, Kraus:2001id, Kraus:2002nu}, and its relation to 
the running of the holomorphic gauge coupling was later elucidated in Ref.~\cite{Babington:2005vu}. This shows that the quantum effective action \eqref{eq:L:superspace} does not possess the classical conformal invariance of  the same theory with constant couplings. In other words, the trace anomaly and the existence of a renormalization group flow is encoded in the quantum effective action at the classical level already.

In the absence of a conserved supercurrent, the definition of the energy-momentum tensor becomes ambiguous, leading in particular to different interpretations of its trace in different renormalization schemes~\cite{Shifman:1986zi, ArkaniHamed:1997mj}.
In our case, the situation only seems worse with the existence of a dimensionless superfield $G$ which allows for various improvement terms for the energy-momentum tensor, leading to a serious ambiguity in the definition of the classical energy density.
We will see in the next section that the solution to this puzzle is to impose an equation of motion for the coupling superfield $G$, as in eq.~\eqref{eq:eom:teaser}.

\subsection{The spontaneous breaking of conformal symmetry}

In the example of a pure $\N = 1$ gauge theory discussed above, eq.~\eqref{eq:J:nonconservation} can be rewritten as
\begin{equation}
	\D^\alpha J_{\alpha\alphadot}
	\propto \big( \bar{\D}_\alphadot \bar{G} \big) \,
	\frac{\delta \Gamma}{\delta \bar{G}}.
\end{equation}
This relation persists when chiral superfields are added to the theory: as shown in Appendix~\ref{sec:supercurrent}, there exist a unique supercurrent whose divergence is proportional to the variation of the action with respect to the coupling superfields. 
This observation has a straightforward interpretation: if $G$ and $\bar{G}$ are interpreted as quantum fields, that is if the path integral is performed over them (or equivalently over the cutoff superfield $\Lambda$), then the theory is conformally invariant:
indeed, if the theory obeys the superfields equations of motion,
\begin{equation}
	\frac{\delta \Gamma}{\delta G} = 0,
	\hspace{1.5cm}
	\frac{\delta \Gamma}{\delta \bar{G}} = 0,
	\label{eq:eom:G:principle}
\end{equation}
the energy-momentum tensor is part of a divergenceless current superfield, and its trace is identically zero. Contrarily to the theory with constant couplings, no new counterterms are needed for composite operators, and it can be expected that this conformal symmetry is preserved in the full quantum theory.
There is still an apparent breaking of scale invariance due to the non-trivial transformation of the path integral measure~\cite{Fujikawa:1979ay}, but it can be compensated by a shift symmetry in the coupling, i.e.~the quantum theory is invariant under the combined transformation
\begin{equation}
	x^\mu \to e^\sigma \, x^\mu,
	\hspace{1.5cm}
	\phi(x) \to e^{-\sigma} \, \phi(e^\sigma x)
	\hspace{1.5cm}
	G \to G + \sigma.
	\label{eq:shiftsymmetry}
\end{equation}
Conformal invariance must eventually be broken by the vacuum of the theory, leading to the known behavior of the $\N = 2$ theory, that of a theory with an implicit scale dependence. Such a scenario is not new, but has been conjectured and studied in multiple occasions in the literature, with potential consequences for naturalness and the cosmological constant problem~\cite{Fubini:1976jm, Antoniadis:1984kd, Wetterich:1987fm, Shaposhnikov:2008xi, Englert:2013gz, Hambye:2013sna, Tamarit:2013vda, Salvio:2014soa}. Here, we provide an explicit realization of such a spontaneous conformal symmetry breaking scenario in an $\N = 2$ gauge theory as a soft deformation of the $\N = 4$ supersymmetric Yang-Mills theory.
Note that the (derivative) contact interaction for the current superfield $G$ in the action~\eqref{eq:L:superspace} play a crucial role here: without them, imposing the equation of motion~\eqref{eq:eom:G:principle} is equivalent to setting all interaction terms to zero, meaning that the quantum theory is forced to be trivially free; with the presence of such terms, the equations of motion for the coupling fields are non-trivial and result in new stationary solution that are studied in the next section.

The informed reader will certainly have realized at this point that this interpretation of the theory is not as compelling as is might seen at first.
Besides allowing for interesting solitonic solution, the new terms in the quantum effective action bring in trouble: as best seen in the action~\eqref{eq:L} in components, these terms contain four derivatives of the field $g$, and theories of this type have been known for a very long time to be ill-defined. The energy is unbounded below already at the semi-classical level, as will be seen in the computation of the energy in Section~\ref{sec:energy}, and this prevents any sensible quantization without negative-norm ``ghost'' states~\cite{Nicolis:2008in, Nicolis:2009qm}. 
Yet, it should be mentioned that higher-derivative theories of the type encountered here are the subject of a number of studies in the literature, and that there might be ways of making sense of them as complete quantum theories (see e.g.~Refs.~\cite{Lee:1969fy, Anselmi:2017lia, Salvio:2014soa}):
the central idea behind these approaches is that the negative-norm states do not belong to the ``physical'' Hilbert space, in a similar fashion as in the BRST quantization of gauge theories in which Faddeev-Popov ghost never appear as physical states.%
\footnote{This possibility is somehow supported by the axiomatic conformal field theory (CFT) point of view, where the problem of ghosts is related to the existence of a scalar field $g$ with scaling dimension zero, breaking the unitarity bound for CFT operators:
due to the shift symmetry~\eqref{eq:shiftsymmetry}, however, the field $g$ cannot be a primary operator of the CFT.}
Moreover, as serious as this problem may seem, one should keep in mind that the action~\eqref{eq:L:superspace:cutoff} is computed in an effective theory framework, integrating out part of the $\N = 4$ superfield content below the cutoff scale $\langle \Lambda(x) \rangle$, and using moreover a derivative expansion for the field $\Lambda$. The unstable modes that appear due to the presence of four-derivative terms are actually outside the realm of this effective approach, and one can expect that the UV completion of the model consisting in an $\N = 4$ theory deformed with a soft mass term is free of such ghosts states. This is an interesting question that is left open for future studies. For now, Section~\ref{sec:semiclassics} will focus on the study of static solutions (i.e.~solution with no time dependence), for which the problem of ghosts is absent, as explicitly verified below.

Let us finally remark that there exist a different interpretation of the theory that avoids this delicate problem, if $G$ is taken to be an auxiliary (unquantized) field. The computation of physical observables only becomes independent of the cutoff when one includes all higher-order corrections in inverse powers of the cutoff and its derivatives. In particular, results obtained in a semi-classical approach depend on the choice of $G$. 
Yet, in a search for extremal configuration of the action, it is natural to impose the condition~\eqref{eq:eom:G:principle}. 
It appears therefore that imposing the equation of motion for the coupling field is in any case the only way to make sense of the quantum effective action~\eqref{eq:L} semi-classically. Our results will therefore be valid in both interpretations, even though they are two distinct quantum field theories:
if the superfield $G$ is dynamical, there is an additional degree of freedom that can be interpreted as a dilaton.%
\footnote{The dilaton is the would-be Goldstone boson of the spontaneously-broken dilatation invariance~\cite{Nielsen:1975hm, Low:2001bw}, which is however not expected to arise as a massless mode in the effective action~\cite{Holdom:1986ub, Holdom:1987yu}, excepted in some very special cases~\cite{Coradeschi:2013gda, Bellazzini:2013fga}.}
We leave also open a number of questions regarding the interpretation of our theory, and proceed with the semi-classical analysis that leads to the appearance of a massless monopole solution.


\section{The semi-classical approximation and a massless monopole}
\label{sec:semiclassics}

The starting point of the semi-classical analysis of the action~\eqref{eq:L} is the equations of motion for the gauge and matter fields, which get modified in the presence of a spacetime-dependent coupling,
\begin{equation}
	D^\nu F_{\mu\nu}
	+ \frac{i}{2} \, \phi^\dag \, \overleftrightarrow{D_\mu} \phi
	+ \frac{i}{2} \, \phi \, \overleftrightarrow{D_\mu} \phi^\dag
	= - \frac{\partial^\nu g}{g} \, F_{\mu\nu}
	\label{eq:eom:A}
\end{equation}
and
\begin{equation}
	D_\mu D^\mu \phi
	+ \frac{1}{2} \, \big[ \phi, [ \phi^\dag, \phi ] \big] =
	- \frac{\partial_\mu g}{g} \, D^\mu \phi
	- \frac{1}{2} \, \frac{\square g}{g} \, \phi.
	\label{eq:eom:phi}
\end{equation}
In both equations, the left-hand side corresponds to the ordinary equation of motion, vanishing at constant coupling, but the right-hand side is new.
In addition we impose the equation of motion for the coupling $g$,
\begin{eqnarray}
	\frac{1}{2} \, \tr\big( F_{\mu\nu} F^{\mu\nu} \big)
	+ \frac{1}{2} \, \tr\big( \phi^\dag D_\mu D^\mu \phi \big)
	+ \frac{1}{2} \, \tr\big( \phi \, D_\mu D^\mu \phi^\dag \big)
	+ \frac{1}{4} \, \tr\big( [ \phi^\dag, \phi ]^2 \big)
	\nonumber \\
	= \frac{c}{b} \left[ \square^2 g
	- \frac{2}{3} \partial_\mu \left( \partial^\mu g \,
	\partial_\nu g \, \partial^\nu g \right) \right].
	\label{eq:eom:g}
\end{eqnarray}
In this case, the vanishing of the left-hand side at constant coupling implies that the action itself must vanish, which is a non-trivial condition for the existence of solitons; we will return to this observation below. The last ingredient needed in the semiclassical approach is the energy-momentum tensor. As mentioned in the previous section, there is a unique conserved supercurrent, constructed explicitly in Appendix~\ref{sec:supercurrent}. Its $\theta\bar{\theta}$ component is the energy-momentum tensor
\begin{eqnarray}
	T^{\mu\nu} & = & b \, \bigg[ \frac{1}{2} \, \eta^{\mu\nu} \,
	g \, \tr\big( F_{\rho\sigma} F^{\rho\sigma} \big)
	- 2 \, g \, \tr\big( F^\mu_{~\rho} F^{\nu\rho} \big)	
	- \frac{1}{12} \, \eta^{\mu\nu} \, g \,
	\tr\big( [ \phi^\dag, \phi ]^2 \big)
	\nonumber \\
	&& \quad
	- \frac{1}{3} \, \eta^{\mu\nu} \, g \, 
	\tr\big( D_\rho \phi^\dag \, D^\rho \phi \big)
	+ \frac{2}{3} \, g \,
	\tr\big( D^\mu \phi^\dag \, D^\nu \phi \big)
	+ \frac{2}{3} \, g \,
	\tr\big( D^\nu \phi^\dag \, D^\mu \phi \big)
	\nonumber \\
	&& \quad
	- \frac{1}{3} \, g \,
	\tr\big( \phi^\dag \, D^\mu D^\nu \phi \big)
	- \frac{1}{3} \, g \,
	\tr\big( \phi \, D^\mu D^\nu \phi^\dag \big)
	- \frac{1}{3} \, \partial^\mu \partial^\nu g \,
	\tr\big( \phi^\dag \phi \big)
	\nonumber \\
	&& \quad
	- \frac{1}{6} \, \eta^{\mu\nu} \, \partial_\rho g \,
	\tr\big( \phi^\dag D^\rho \phi \big)
	+ \frac{1}{6} \, \partial^\mu g \,
	\tr\big( \phi^\dag D^\nu \phi \big)
	+ \frac{1}{6} \, \partial^\nu g \,
	\tr\big( \phi^\dag D^\mu \phi \big)
	\nonumber \\
	&& \quad
	- \frac{1}{6} \, \eta^{\mu\nu} \, \partial_\rho g \,
	\tr\big( \phi \, D^\rho \phi^\dag \big)
	+ \frac{1}{6} \, \partial^\mu g \,
	\tr\big( \phi \, D^\nu \phi^\dag \big)
	+ \frac{1}{6} \, \partial^\nu g \,
	\tr\big( \phi \, D^\mu \phi^\dag \big) \bigg]
	\nonumber \\
	&& + c \, \bigg[
	- \frac{1}{2} \, \eta^{\mu\nu} \, \square g \, \square g
	+ 2 \, \partial^\mu \partial^\nu g \, \square g
	+ \frac{1}{3} \, \eta^{\mu\nu} \, \partial_\rho \partial_\sigma g \,
	\partial^\rho \partial^\sigma g
	- \frac{4}{3} \, \partial^\mu \partial_\rho g \,
	\partial^\nu \partial^\rho g
	\nonumber \\
	&& \qquad
	+ \frac{1}{3} \, \eta^{\mu\nu} \, \partial_\rho g \,
	\partial^\rho \square g
	- \partial^\mu g \, \partial^\nu \square g
	- \partial^\nu g \, \partial^\mu \square g
	+ \frac{2}{3} \, \partial_\rho g \,
	\partial^\mu \partial^\nu \partial^\rho g
	\nonumber \\
	&& \qquad
	- \frac{1}{6} \, \eta^{\mu\nu} \,
	\left( \partial_\rho g \, \partial^\rho g \right)^2
	+ \frac{2}{3} \, \partial^\mu g \, \partial^\nu g \,
	\partial_\rho g \, \partial^\rho g
	\bigg]
	+ \ldots
	\label{eq:T}
\end{eqnarray}
where we have only kept the lowest component $g$ of the coupling superfield $G$ and the bosonic components of the gauge and matter superfields, as in the action~\eqref{eq:L}.
Ignoring the fermions is a choice that simplifies vastly the search for solitons, and is can be justified by noting that all the equations of motion for fermionic fields are satisfied when the latter are set to zero. There might exist other semi-classical solutions to the theory that include non-zero fermionic fields, but we simply ignore this possibility for now and leave it open to further studies.
Appendix~\ref{sec:supercurrent} contains an explicit expression for the energy-momentum tensor with the bosonic superpartners of the coupling $g$, see eq.~\eqref{eq:T:complete}, but they do not play a role in the description of the massless monopole solution.

The fact that the energy-momentum tensor~\eqref{eq:T} is unique is really a consequence of using a spacetime-dependent coupling and imposing an equation of motion for it: in a general supersymmetric theory, the ordinary Ferrara-Zumino multiplet~\cite{Ferrara:1974pz}, can be ``improved'' with various terms, leading to a some ambiguity in the definition of the supercurrent~\cite{Komargodski:2010rb, Nakayama:2012nd}; here there is a unique superfield that satisfies eq.~\eqref{eq:supercurrent:conservation}.

In summary, the rules of the game for a semi-classical analysis of this model are simply to find solutions to the equations of motion~(\ref{eq:eom:A}--\ref{eq:eom:g}) and study their energy given by the component $T^{00}$ of eq.~\eqref{eq:T}.

\subsection{Boundedness of the energy}
\label{sec:energy}

A pivotal result in the study of $\N = 2$ supersymmetric theories was the discovery of a topological bound on the energy of monopoles and dyons~\cite{Bogomolny:1975de, Faddeev:1976pg}, and of its relation to the central charge of the $\N = 2$ supersymmetry algebra~\cite{Witten:1978mh}. Setting the coupling $g$ to a constant value and making use of the equations of motion, the energy of a field configuration $E = \int \d^3x \, T^{00}$ obtained from the energy-momentum tensor~\eqref{eq:T} can indeed be written as a sum of positive terms and boundary terms. These terms can moreover be combined in such a way that the total energy satisfies the known topological bound.

With non-constant couplings, it seems that there is unfortunately no obvious way of writing the energy in the form of a manifestly positive quantity.%
\footnote{It should be repeated here that our formulation~\eqref{eq:L:superspace:cutoff} of the theory explicitly breaks one of the supersymmetries by introducing a cutoff that only transform as a $\N = 1$ superfield. It would be interesting to study the energy-momentum tensor in an explicitly $\N = 2$ invariant formulation to see if a relation can be drawn between the semi-classical energy and the central charge of the supersymmetry algebra.
}
It is not obvious either that solutions to the equations of motion with negative infinite energy exist, however. We leave the general problem open for future investigations, and focus on a particular situation of physical interest: we consider static solutions, i.e.~time-independent ones, and work in the temporal gauge $A_0 = 0$. In this case, the energy density becomes
\begin{eqnarray}
	T^{00} & = & b \, \bigg[ \frac{1}{2} \,
	g \, \tr\big( F_{ij} F_{ij} \big)	
	- \frac{1}{12} \, g \, \tr\big( [ \phi^\dag, \phi ]^2 \big)
	\nonumber \\
	&& \quad
	+ \frac{1}{3} \, g \, 
	\tr\big( D_i \phi^\dag \, D_i \phi \big)
	+ \frac{1}{6} \, \partial_i g \,
	\tr\big( \phi^\dag D_i \phi \big)
	+ \frac{1}{6} \, \partial_i g \,
	\tr\big( \phi \, D_i \phi^\dag \big) \bigg]
	\nonumber \\
	&& + c \, \bigg[
	- \frac{1}{2} \, \left( \partial_i^2 g \right)^2
	+ \frac{1}{3} \, \left( \partial_i \partial_j g \right)^2
	+ \frac{1}{3} \, \partial_i g \, \partial_i \partial_j^2 g
	- \frac{1}{6} \, \left( \partial_i g \, \partial_i g \right)^2 \bigg],
	\label{eq:T00:static}
\end{eqnarray}
where $i = 1, 2, 3$ runs over spatial indices only. Substituting the equations of motion \eqref{eq:eom:phi} and \eqref{eq:eom:g}, this can equivalently be rewritten as 
\begin{equation}
	T^{00} = c \, \bigg[
	\frac{1}{2} \, \left( \partial_i^2 g \right)^2
	+ \frac{1}{2} \, \left( \partial_i g \, \partial_i g \right)^2 \bigg]
	+ \partial_i \mathcal{B}_i,
	\label{eq:T00:static:simplified}
\end{equation}
namely in the form of positive terms and a boundary term $\partial_i \mathcal{B}_i$ in which
\begin{eqnarray}
	\mathcal{B}_i & = & b \, \bigg[
	\frac{1}{6} \, g \, \partial_i \tr\big( \phi^\dag \phi \big)
	- \frac{1}{3} \, \partial_i g \, \tr\big( \phi^\dag \phi \big) \bigg]
	\nonumber \\
	&& + c \, \bigg[
	g \, \partial_i \partial_j^2 g
	- \partial_i g \, \partial_j^2 g
	+ \frac{1}{3} \, \partial_j g \, \partial_i \partial_j g
	- \frac{2}{3} \, g \, \partial_i g \, \left( \partial_j g \right)^2 \bigg].
\end{eqnarray}
With suitable boundary conditions at spatial infinity, namely the vanishing of derivatives of $g$ and $\tr( \phi^\dag \phi )$, the total energy of a static configuration is non-negative.
Moreover, the bound can only be saturated if $g$ is strictly constant. In this case, the equation of motion \eqref{eq:eom:g} requires the action to vanish identically, and the only solutions are of the vacuum type, in which $\phi = 0$ and $A_\mu$ is in a pure gauge configuration.

This results motivates the search for non-trivial solutions to the equations of motion, necessarily including a non-constant coupling field. This task is the object of the next section, resulting in the observation of a new type of soliton.
Before moving on, let us specify that the analysis of the boundedness of the energy of a static configuration is repeated in Appendix~\ref{sec:supercurrent} including the bosonic superpartners of the gauge coupling $g$, and that it raises an issue: 
the coupling superfield $G$ has a $\theta^2$ component denoted by $f$ that is not an auxiliary field, but possesses the kinetic term of an ordinary complex field due to the presence of higher-derivative interactions in the action~\ref{eq:L:superspace}; it turns out that the potential for $f$ is unbounded, as it is dominated at large $f$ by a quartic term with the wrong sign. However, the trivial solution $f = 0$ to the equations of motion is still a local minimum of the energy, around which it makes sense to search for solitons.

\subsection{Spherically-symmetric solitons and the massless Wu-Yang monopole}

The strategy to search for solitons is usually to start with an ansatz with the largest possible symmetry. Using Lorentz-invariance of the action, one can look for a soliton at rest, recovering the assumption of time-independence made in the previous section, and moreover invariant under $SO(3)$ rotational symmetry combined with the internal gauge symmetry. Such a spherically-symmetric ansatz can be obtained taking for instance~\cite{Goddard:1977da}
\begin{equation}
	A^i = \epsilon^{ijk} \left[ 1 - f(r) \right] \frac{x^j}{r^2} \, T^k,
	\hspace{1.5cm}
	\phi = h(r) \, \frac{x^i}{r^2} \, T^i,
	\hspace{1.5cm}
	g = g(r),
	\label{eq:sphericallysymmetricansatz}
\end{equation}
where $f(r)$, $h(r)$ and $g(r)$ are functions of the radial direction $r^2 = x_i^2$ to be determined, and the $T^i$ are generators of some $SU(2)$ subgroup of the full gauge group. The equations of motion are then reduced to ordinary differential equations for the radial functions, of which we can hope to find a solution.
A famous configuration described by this ansatz is the BPS monopole~\cite{Prasad:1975kr}, with
\begin{equation}
	f(r) = \frac{v \, r}{\sinh(v \, r)},
	\hspace{1.5cm}
	h(r) = \frac{v \, r}{\tanh(v \, r)} - 1,
	\label{eq:BPS}
\end{equation}
This is a solution of eqs.~\eqref{eq:eom:A} and \eqref{eq:eom:phi} for constant coupling $g(r) = \text{const}$. At large distances from the origin $h(r) \to v r$, meaning that $\phi$ has a non-trivial vacuum expectation value $\tr(\phi^\dag \phi) = \frac{1}{2} v^2$. In the same limit $f(r) \to 0$ and the gauge field approaches a magnetic monopole configuration.
What is peculiar to our approach is that the BPS monopole configuration~\eqref{eq:BPS} with constant coupling is not a solution of the equation of motion~\eqref{eq:eom:g} for $g$. Thus it does not qualify as a semi-classical solution of the $\N = 2$ supersymmetric Yang-Mills theory with spacetime-dependent couplings.
We will actually see in the next section that the ordinary BPS monopole still arises as a semi-classical solution in this model, but only in the limit $N_c \to \infty$. At finite $N_c$, it must be discarded.

Another historical solution to the Yang-Mills equation of motion falls into the ansatz of eq.~\eqref{eq:sphericallysymmetricansatz}: it is the Wu-Yang monopole~\cite{Wu:1975vq}, corresponding to $f(r) = h(r) = 0$. In that case, since the field $\phi$ is identically zero, its equation of motion~\eqref{eq:eom:phi} is automatically satisfied. On the other hand, the gauge field strength tensor is non-zero for such a solution, but eq.~\eqref{eq:eom:A} is nevertheless satisfied when the coupling is an arbitrary function of the radius due to the property $x_i F_{ij} = 0$. The remaining equation of motion for $g(r)$ can then be solved, and we have the solution
\begin{equation}
	A^i(x) = \epsilon^{ijk} \, \frac{x^j}{r^2} \, T^k,
	\hspace{1.5cm}
	\phi(x) = 0,
	\hspace{1.5cm}
	g(x) = \omega \, \log\left( \frac{r}{r_0} \right)
	\label{eq:WuYang}
\end{equation}
where $r_0$ is an arbitrary length and $\omega$ satisfies the cubic equation $\omega (\omega^2 + 3) = 3 b/ 4 c$, i.e.
\begin{equation}
	\omega = \left[ \sqrt{1 + \left( \frac{3 \, b}{8 \, c} \right)^2}
	-  \frac{3 \, b}{8 \, c} \right]^{-1/3}
	- \left[ \sqrt{1 + \left( \frac{3 \, b}{8 \, c} \right)^2}
	-  \frac{3 \, b}{8 \, c} \right]^{1/3}.
	\label{eq:omega}
\end{equation}
For any $SU(N_c)$ gauge theory with $N_c \geq 2$, $\omega$ is a real, positive number contained in the interval $(0,1)$.
The existence of such a solution to the equations of motion is remarkable, but it is not by itself guaranteed to be of physical relevance. There is indeed a serious problem with the Wu-Yang solution: it is singular at the origin.
With constant coupling, the gauge part of energy density obeys
\begin{equation}
	T^{00} \sim \frac{1}{2} \, \tr\big( F_{ij} F_{ij} \big) = \frac{1}{2 r^4}.
\end{equation}
This singularity is not integrable, and the total energy of an ordinary Wu-Yang monopole is therefore divergent.
In our approach, the logarithmic dependence of the coupling on the radial direction does not remove the singularity at $r = 0$, but allows to regularize the energy.
Plugging the solution~\eqref{eq:WuYang} into the energy density~\eqref{eq:T00:static}, we obtain
\begin{equation}
	T^{00} = \frac{\omega \, b}{2 \, r^4} \,
	\left[ \log\left( \frac{r}{r_0} \right) - \delta \right],
	\hspace{1.5cm}
	\delta = \frac{c}{3 b} \, \omega \, \left( 1 + \omega^2 \right).
\end{equation}
$\delta$ is a number between $\frac{1}{8}$ and $\frac{1}{12}$ depending on the gauge group, and accounts for the negative contribution to the energy of the terms with derivatives of $g$.
Introducing a regulator $\epsilon$ at short distances from the origin, a regularized version of the energy can be defined as
\begin{equation}
	E = \pi \, \omega \, b \, 
	\int_\epsilon^\infty \d r \, \frac{\log(r/r_0) - \delta}{r^2}
	\xrightarrow{\epsilon \to 0} 0
	\hspace{1cm}
	\text{if}~r_0 = \epsilon \, e^{1-\delta}.
\end{equation}
As indicated, the divergence when $\epsilon \to 0$ cancels provided that we precisely adjust the parameter $r_0$ of the solution to be proportional to $\epsilon$.
The use of a regulator can seem to be \emph{ad hoc}, but it makes sense in the effective theory framework of Section~\ref{sec:effectiveaction}: after all, the quantum effective action that we derived in eq.~\eqref{eq:L:superspace:cutoff} is only valid at energy scales below $\Lambda$, i.e.~it breaks down at very short distances.

The Wu-Yang solution is called a monopole since at large distances from the origin, the gauge potential mimics the presence of a magnetic charge.
While the energy density is positive and approaching zero far from the monopole, it is large and negative close to the regulator scale $\epsilon$. This results in a soliton with zero overall mass, but with a divergent radius,
\begin{equation}
	\langle r \rangle = \int \d^3x \, r \, T^{00} \to \infty.
\end{equation}
This behavior could have been expected in a theory that does not break scale invariance.
From this point of view, the logarithmic dependence of the gauge coupling $g$ on the radius $r$ is satisfactory: it is reminiscent of the logarithmic dependence of the holomorphic gauge coupling on the renormalization scale in a traditional approach. Our solution permits to identify the renormalization scale with the inverse of the distance to the magnetic charge, the proportionality factor being fixed by the quantity $\omega$.
What looks slightly disturbing when taking a closer look at the solution~\eqref{eq:WuYang} is that the holomorphic coupling $g$ turns negative in a region comprised between $\epsilon$ and $r_0$. This is actually not in contradiction with our definition~\eqref{eq:G:def} of the coupling in terms of the UV cutoff of the theory, but it indicates that the effective theory approach breaks down in that region.

Other solitons could actually exist in this system, for instance with non-zero vacuum expectation value for the field $\phi$, but an exhaustive search goes beyond the scope of this work. If such a general, regular solutions exist, one can hope to recover the solution~\eqref{eq:WuYang} is some limit, hence gaining insight on its physical interpretation. 
For now, we simply mention one framework in which exact, semi-classical solutions can be found more easily: the limit $N_c \to \infty$.

\subsection{Large $N_c$ limit}

Spherically-symmetric solitons obtained using an ansatz of the type of eq.~\eqref{eq:sphericallysymmetricansatz} only rely on the existence of a $SU(2)$ subgroup of the gauge group of the theory, and from this point of view their properties should not depend on the number of colors $N_c$. There is however an explicit appearance of group-theoretical factors in the equation of motion~\eqref{eq:eom:g}, through the relative dependence of the coefficients $b$ and $c$ on $N_c$. This is visible in the Wu-Yang monopole solution of the last section, in which the factor $\omega$ depends on $N_c$; it actually vanishes when $N_c \to \infty$, and at leading non-zero order, 
\begin{equation}
	\omega = \frac{3}{2} \, N_c^{-1}
	+ \O\left( N_c^{-2} \right).
\end{equation}
At leading order in $N_c$, the action is indeed dominated by the vacuum terms proportional to $c$, and it can only be stationary with constant coupling. The equation of motion \eqref{eq:eom:g} is trivially satisfied by a constant $g$ at order $N_c^2$. The problem of finding solitons at large $N_c$ reduces therefore to the ordinary case of constant couplings, meaning that the BPS monopoles described by eqs.~\eqref{eq:sphericallysymmetricansatz} and \eqref{eq:BPS} are actual solutions.
This is in agreement with the common lore that the large $N_c$ limits of gauge theories are somehow ``classical''.

This observation allows to outline a strategy for finding solitons at finite $N_c$, working order by order. The first correction to the ordinary BPS solution arises at order $1/N_c$ for all the fields: the correction to $g$ can be readily extracted from eq.~\eqref{eq:eom:g}, and then plugged into eqs.~\eqref{eq:eom:A} and \eqref{eq:eom:phi} to find the corrections to $\phi$ and $A_\mu$.
At the level of the energy density, these corrections are of order $N_c^0$, as opposed to the classical BPS monopole mass that is of order $N_c$ in our normalization.
The procedure can be iterated at all orders in $N_c$, at least in principle, and one can possibly uncover new solitonic solutions in this way.
Even though the method is straightforward, the task is rather involved and we leave it for future work.


\section{Conclusions}
\label{sec:conclusion}

In summary, this work suggests a new approach to the search for solitons in quantum field theory, in which coupling constants are replaced by expectation values of fields, not necessarily constant over space and time.
The method allows to find solitons in theories that do not have dimensionful quantities in their Lagrangian. So far, all known types of solitons appearing in four-dimensional quantum field theories were linked to a specific mass scale: the BPS monopoles and dyons have masses proportional to the vacuum expectation value of the Higgs field; skyrmions in the chiral Lagrangian of QCD have an energy directly related to the pion decay constant. In an effective action that is classically scale invariant, it seems that nothing can stabilize the mass of a soliton. We showed that it is possible to obtain non-trivial solitons by simply requiring stationarity of the action with respect to the coupling field.%
\footnote{This requirement is close in spirit to Derrick's theorem~\cite{Derrick:1964ww} and its relatives (see e.g.~Ref.~\cite{Manton:2008ca}), stating that the energy of a soliton is stationary with respect to a uniform spatial rescaling, an argument usually used to exclude the presence of solitons in a wide range of systems.}
In a sense, this accounts for finding an optimal cutoff scale at each point in space and time.

This feature is only possible thanks to the presence of terms containing derivatives of the coupling fields in the action.
In this work, we used a soft deformation of the conformal $\N = 4$ supersymmetric Yang-Mills theory to obtain the quantum effective action of a $\N = 2$ theory in a one-loop computation.
Effective action terms involving derivatives of the couplings are however present in any theory, as soon as one requires the action to generate finite correlators for composite operators. It would be particularly interesting but also very challenging to perform a search for solitons in a theory without any supersymmetry such as QCD, where higher-loop corrections must be taken into account in the effective action.

We have also found in this paper a new realization of the old Wu-Yang monopole, with a better behaviour than its original cousin, despite leaving a few questions unanswered. In future work, one can hope to find a soliton that interpolates between the BPS monopole at large $N_c$ and finite Higgs vacuum expectation value, and the Wu-Yang solution at finite $N_c$ and zero Higgs field, with the hope to reproduce the known behavior of $\N = 2$ gauge theories in a purely semi-classical approach. If the Wu-Yang monopole can indeed be interpreted as a true massless state of the theory, it could represent the S-dual in the IR to the gluon of an unbroken gauge theory in the UV~\cite{Seiberg:1994rs}.
In this case, the vacuum of the theory could be realized as a superposition of massless monopole states.

Among the numerous interesting questions that can be addressed by extending the present work is also the strong CP problem and the physics of the $\theta$-angle, which we have set to zero in our analysis. As mentioned above, a complete $\N = 2$ formulation of the quantum effective action would be required to do so.
Notice also that while supersymmetry forbids the presence of a cosmological constant in the action, the vacuum can obtain a finite --- and even negative --- energy if couplings vary over space and time, with potentially intriguing consequences for cosmology.

\acknowledgments

This project is supported by the Swiss National Science Foundation (SNSF) through the NCCR SwissMAP and also under grant no.~P300P2154559.


\appendix

\section{Supercurrent and energy-momentum tensor}
\label{sec:supercurrent}

This appendix contains the results of the computation of the supercurrent, first derived in the superspace formalism, and then explicitly given in components for its bosonic part. The superspace notation used throughout the paper follows from the definition of the supersymmetric covariant derivatives
\begin{equation}
	\D_\alpha = \frac{\partial}{\partial \theta^\alpha}
	+ i \, \bar{\theta}^\alphadot \left( \sigma^\mu \right)_{\alpha\alphadot}
	\partial_\mu ,
	\hspace{1.5cm}
	\bar{\D}_\alphadot = -\frac{\partial}{\partial \bar{\theta}^\alphadot}
	- i \, \theta^\alpha \left( \sigma^\mu \right)_{\alpha\alphadot} \partial_\mu ,
	\label{eq:covariantderivatives}
\end{equation}
where $\sigma^\mu$ are the usual Pauli matrices and $\theta^\alpha$ and $\bar{\theta}^\alphadot$  the anticommuting superspace coordinates. 
We also use the conventional shorthand notation $\D^2 = \D^\alpha \D_\alpha$ and $\bar{\D}^2 = \bar{\D}_\alphadot \bar{\D}^\alphadot$, and similarly for the contraction of spinor fields. The superfields $\Phi$ and $G$ are chiral, $\bar{\Phi}$ and $\bar{G}$ antichiral, that is they satisfy $\bar{D}_\alphadot \Phi = \bar{D}_\alphadot G = D_\alpha \bar{\Phi} = D_\alpha \bar{G} = 0$, and $V = V^\dag$ is a vector superfield. The theory is invariant under the non-linear gauge transformation 
\begin{equation}
	\Phi \to e^{-\Omega} \Phi e^{\Omega},
	\qquad
	\bar{\Phi} \to e^{-\bar{\Omega}} \bar{\Phi} e^{\bar{\Omega}},
	\qquad
	e^{2V} \to e^{-\Omega} e^{2V} e^{\bar{\Omega}},
	\label{eq:gaugetransformation}
\end{equation}
where $\Omega$ and $\bar{\Omega}$ are respectively chiral and antichiral superfields.
For quantities that are not gauge singlets, we also use the gauge-covariant derivatives $\covD_\alpha$ and $\bar{\covD}_\alphadot$ instead of \eqref{eq:covariantderivatives}, satisfying for instance
\begin{equation}
	\covD_\alpha \Phi =
	e^{2V} \D_\alpha \left( e^{-2V} \Phi e^{2V} \right) e^{-2V}.
	\label{eq:gaugecovariantderivative:Phi}
\end{equation}
The holomorphic and anti-holomorphic gauge field strength tensors are defined as
\begin{equation}
	\W_\alpha = -\frac{1}{8} \, \bar{\D}^2
	\left( e^{2 V} \D_\alpha e^{-2V} \right),
	\hspace{1cm}
	\bar{\W}_\alphadot = -\frac{1}{8} \, \D^2
	\left( e^{-2 V} \bar{\D}_\alphadot e^{2V} \right),
	\label{eq:W}
\end{equation}
and they satisfy
\begin{equation}
	\covD_\alpha \bar{\W}_\alphadot = \bar{\covD}_\alphadot \W_\alpha = 0,
	\hspace{1.5cm}
	\covD^\alpha \W_\alpha = e^{2V} \,
	\big( \bar{\covD}_\alphadot \bar{\W}^\alphadot \big) \, e^{-2V}.
	\label{eq:W:properties}
\end{equation}

\subsection{The equations of motion in superspace}

The equations of motion can be derived directly in superspace from the Lagrangian \eqref{eq:L:superspace}. The conservation of the supercurrent follows from these superfield equations, so it is useful to write them down explicitly.
For the coupling superfields, we have
\begin{eqnarray}
	\frac{\delta S}{\delta \bar{G}}
	& = & 2 \sqrt{2} \, b \, 
	\tr\big( \bar{\W}_\alphadot \bar{\W}^\alphadot \big)
	\nonumber \\
	&& - \frac{1}{4} \, \D^2 \left[ \frac{b}{\sqrt{2}} \,
	\tr\big( \bar{\Phi} e^{-2V} \Phi e^{2V} \big)
	- c \, \square G
	- \frac{c}{12} \, \bar{\D}_\alphadot
	\left( \D^\alpha G \, \D_\alpha G \, \bar{\D}^\alphadot \bar{G} \right)
	\right],
	\label{eq:superfieldeq:Gbar}
\end{eqnarray}
and for the matter superfield
\begin{eqnarray}
	\frac{\delta S}{\delta \bar{\Phi}}
	& = & -b \, \frac{1}{4} \, \D^2
	\left[ \frac{1}{\sqrt{2}} \left( G + \bar{G} \right) 
	e^{-2V} \Phi e^{2V} \right],
	\label{eq:superfieldeq:Phibar}
\end{eqnarray}
as well as the complex conjugate of these two equations.
The superspace equation of motion for $V$ can be written as
\begin{eqnarray}
	\frac{\delta S}{\delta V}
	& \equiv &
	b \, \bigg[ - \sqrt{2} \, e^{-V} \mathscr{D}^\alpha
	\left( G \, \W_\alpha\right) e^V
	- \sqrt{2} \, e^V \bar{\mathscr{D}}_\alphadot
	\left( \bar{G} \, \bar{\W}^\alphadot \right) e^{-V}
	\nonumber \\
	&& \qquad
	+ \frac{1}{\sqrt{2}} \, \left( G + \bar{G} \right) 
	\left( e^{-V} \Phi e^{2V} \bar{\Phi} e^{-V}
	+ e^V \bar{\Phi} e^{-2V} \Phi e^V \right)
	\bigg].
	\label{eq:superfieldeq:V}
\end{eqnarray}
This quantity transforms non-linearly under gauge transformations, but the combinations $e^{-V} \frac{\delta S}{\delta V} e^V$ and $e^V \frac{\delta S}{\delta V} e^{-V}$ both transform covariantly.

\subsection{The supercurrent as a superfield}

There is a unique supercurrent written in terms of $V$, $\Phi$ and $G$ which is conserved by the above equations of motion, up to an overall normalization.
It is
\begin{eqnarray}
	J_{\alpha\alphadot}
	& = & \sqrt{2} \, b \, \bigg[
	\left( G + \bar{G} \right) \tr\big( \W_\alpha e^{2V}
	\bar{\W}_\alphadot e^{-2V} \big)
	+ \frac{1}{48} \, \left( G + \bar{G} \right) \,
	\left[ \D_\alpha, \bar{\D}_\alphadot \right]
	\tr\big( \bar{\Phi} e^{-2V} \Phi e^{2V} \big)
	\nonumber \\
	&& \qquad\quad
	+ \frac{1}{16} \, \left( G + \bar{G} \right)
	\tr\big( \bar{\covD}_\alphadot \bar{\Phi} \, e^{-2V} \,
	\covD_\alpha \Phi \, e^{2V} \big)
	\nonumber \\
	&& \qquad\quad
	- \frac{1}{48} \, \D_\alpha G \, \bar{\D}_\alphadot
	\tr\big( \bar{\Phi} e^{-2V} \Phi e^{2V} \big)
	+ \frac{1}{48} \, \bar{\D}_\alphadot \bar{G} \, \D_\alpha
	\tr\big( \bar{\Phi} e^{-2V} \Phi e^{2V} \big)
	\nonumber \\
	&& \qquad\quad
	- \frac{1}{48} \, \left( \bar{\D}_\alphadot \D_\alpha G
	- \D_\alpha \bar{\D}_\alphadot \bar{G} \right)
	\tr\big( \bar{\Phi} e^{-2V} \Phi e^{2V} \big) \bigg]
	\nonumber \\
	&& + c \, \bigg[
	- \frac{1}{24} \, \square \bar{\D}_\alphadot \bar{G} \, \D_\alpha G
	- \frac{1}{24} \, \bar{\D}_\alphadot \bar{G} \, \square \D_\alpha G
	\nonumber \\
	&& \qquad
	- \frac{1}{24} \, \partial_{\alpha\alphadot} \partial^{\beta\dot{\beta}}
	\bar{\D}_{\dot{\beta}} \bar{G} \, \D_\beta G
	- \frac{1}{24} \, \bar{\D}_{\dot{\beta}} \bar{G}
	\partial_{\alpha\alphadot} \partial^{\beta\dot{\beta}} \D_\beta G
	\nonumber \\
	&& \qquad
	- \frac{i}{6} \, \square \bar{G} \, \partial_{\alpha\alphadot} G
	+ \frac{i}{6} \, \partial_{\alpha\alphadot} \bar{G} \, \square G
	+ \frac{i}{12} \, \partial_{\alpha\alphadot}
	\partial_{\beta\dot{\beta}} \bar{G} \, \partial^{\beta\dot{\beta}} G
	- \frac{i}{12} \, \partial^{\beta\dot{\beta}} \bar{G} \,
	\partial_{\alpha\alphadot} \partial_{\beta\dot{\beta}} G
	\nonumber \\
	&& \qquad
	+ \frac{i}{64} \, \partial_{\alpha\alphadot} \bar{\D}^2 \bar{G} \, \D^2 G
	- \frac{i}{64} \, \bar{\D}^2 \bar{G} \, \partial_{\alpha\alphadot} \D^2 G
	\nonumber \\
	&& \qquad
	+ \frac{1}{12} \, \partial^{\beta\dot{\beta}}
	\bar{\D}_{\dot{\beta}} \bar{G} \, \partial_{\beta\alphadot} \D_\alpha G
	+ \frac{1}{12} \, \partial_{\alpha\dot{\beta}}
	\bar{\D}_\alphadot \bar{G} \, \partial^{\beta\dot{\beta}} \D_\beta G
	\nonumber \\
	&& \qquad
	+ \frac{1}{384} \, \D_\alpha \bar{\D}^2 \bar{G} \,
	\bar{\D}_\alphadot \D^2 G
	- \frac{1}{12} \, \partial_\mu \bar{\D}_\alphadot \bar{G} \,
	\partial^\mu \D_\alpha G
	\nonumber \\
	&& \qquad
	+ \frac{i}{48} \, \partial_{\alpha\dot{\beta}} \bar{\D}^{\dot{\beta}} \bar{G} \,
	\bar{\D}_\alphadot \bar{G} \, \D^\beta G \, \D_\beta G
	+ \frac{i}{48} \, \bar{\D}_{\dot{\beta}} \bar{G} \,
	\bar{\D}^{\dot{\beta}} \bar{G} \, \partial_{\beta\alphadot} \D^\beta G \, \D^\alpha G
	\nonumber \\
	&& \qquad
	- \frac{i}{96} \, \partial_{\alpha\alphadot} \bar{G} \, \bar{\D}^2 \bar{G} \,
	\D^\beta G \, \D_\beta G
	+ \frac{i}{96} \, \bar{\D}_{\dot{\beta}} \bar{G} \,
	\bar{\D}^{\dot{\beta}} \bar{G} \, \partial_{\alpha\alphadot} G \, \D^2 G
	\nonumber \\
	&& \qquad
	- \frac{1}{192} \, \bar{\D}^2 \bar{G} \, \bar{\D}_\alphadot \bar{G} \,
	\D^2 G \, \D_\alpha G
	+ \frac{1}{12} \, \partial_{\alpha\dot{\beta}} \bar{G} \,
	\bar{\D}^{\dot{\beta}} \bar{G} \, \partial_{\beta\alphadot} G \, \D^\beta G
	\nonumber \\
	&& \qquad
	- \frac{1}{12} \, \partial_{\alpha\alphadot} \bar{G} \,
	\bar{\D}^{\dot{\beta}} \bar{G} \, \partial_{\beta\dot{\beta}} G \, \D^\beta G
	- \frac{1}{12} \, \partial_{\beta\dot{\beta}} \bar{G} \,
	\bar{\D}^{\dot{\beta}} \bar{G} \, \partial_{\alpha\alphadot} G \, \D^\beta G
	\bigg].
	\label{eq:supercurrent}
\end{eqnarray}
This supercurrent is gauge-invariant term by term, and obeys
\begin{equation}
	\D^\alpha J_{\alpha\alphadot} =
	- \tr\left( e^V \, \bar{\W}_\alphadot \, e^{-V} \,
	\frac{\delta S}{\delta V} \right)
	- \frac{1}{12} \, \bar{\D}_\alphadot \tr\left( \bar{\Phi} \,
	\frac{\delta S}{\delta \bar{\Phi}} \right)
	+ \frac{1}{4} \, \tr\left( \bar{\mathscr{D}}_\alphadot \bar{\Phi} \,
	\frac{\delta S}{\delta \bar{\Phi}} \right)
	+ \frac{1}{4} \, \bar{\D}_\alphadot \bar{G} \,
	\frac{\delta S}{\delta \bar{G}},
\end{equation}
which also implies conservation of the supercurrent written as a four-vector, $J^\mu = (\bar{\sigma}^\mu)^{\alphadot\alpha} J_{\alpha\alphadot}$, by the equations of motion: $\partial_\mu J^\mu = 0$.

\subsection{The Lagrangian in components}

We provide now a number of results explicitly given in terms of ordinary fields (not superfields). We begin with an explicit expression for the Lagrangian \eqref{eq:L:superspace} written in components. For simplicity of notation, we split the Lagrangian into a part that contains gauge and matter fields and a ``vacuum'' part that only contains the couplings,
\begin{equation}
	\L = b \, \L_\text{matter}
	+ c \, \L_\text{vacuum}
\end{equation}
and we discuss both parts separately. We use the notation $A^\mu$ and $D$ for the bosonic components of the vector superfield $V$ and $\xi$ for its fermionic component in the Wess-Zumino gauge. The chiral superfield $\Phi$ contains the field $\phi$ as its lowest component, its superpartner $\chi$ and the auxiliary field $F$, all of them transforming in the adjoint representation of the gauge group. 
We denote by $\frac{1}{\sqrt{2}} (g \pm i \theta)$ the lowest component of $G$ (respectively $\bar{G}$), by $\zeta$ ($\bar{\zeta}$) its fermionic component and by $f$ ($f^*$) its $\theta^2$ ($\bar{\theta}^2$) component.
Starting with the matter part of the Lagrangian, we have
\begin{eqnarray}
	\L_\text{matter} & = &
	- \frac{1}{2} \, g \, \tr\big( F_{\mu\nu} F^{\mu\nu} \big)
	- i \, g \, \tr\big( \bar{\xi} \, \sigma^\mu \,
	\overleftrightarrow{D_\mu} \xi \big)
	- i \, g \, \tr\big( \bar{\chi} \, \sigma^\mu \,
	\overleftrightarrow{D_\mu} \chi \big)
	\nonumber \\ &&
	- \frac{1}{2} \, g \, \tr\big( \phi^\dag \, D_\mu D^\mu \phi \big)
	- \frac{1}{2} \, g \, \tr\big( \phi \, D_\mu D^\mu  \phi^\dag \big)
	+ g \, \tr\big( D^2 \big)
	+ g \, \tr\big(F^\dag F \big) 
	\nonumber \\ &&
	+ \sqrt{2} \, g \, \tr\big(\phi^\dag \, \xi \, \chi \big)
	+ \sqrt{2} \, g \, \tr\big(\phi \, \bar{\xi} \, \bar{\chi} \big)
	- g \, \tr\big( D \, [ \phi^\dag, \phi ] \big)
	\nonumber \\ &&
	- \frac{1}{2} \, \theta \, \tr\big( F_{\mu\nu} \tilde{F}_{\mu\nu} \big)
	- \theta \, \partial_\mu 
	\left[ \tr\big( \bar{\chi} \, \sigma^\mu \, \chi \big)
	+ \tr\big( \bar{\xi} \, \sigma^\mu \, \xi \big)
	- \frac{i}{2} \,
	\tr\big( \phi^\dag \overleftrightarrow{D^\mu} \phi \big) \right]
	\nonumber \\ &&
	- 2 \sqrt{2} \, i \, \tr\big( F_{\mu\nu} \, \zeta \,
	\sigma^{\mu\nu} \, \xi \big)
	+ 2 \sqrt{2} \, i \, \tr\big( F_{\mu\nu} \, \bar{\zeta} \,
	\bar{\sigma}^{\mu\nu} \, \bar{\xi} \big)
	\nonumber \\ &&
	+ \sqrt{2} \, \tr\big( D \, \zeta \, \xi \big)
	+ \sqrt{2} \, \tr\big( D \, \bar{\zeta} \, \bar{\xi} \big)
	+ \sqrt{2} \, \tr\big( [\phi^\dag, \phi ] \, \zeta \, \xi \big)
	- \sqrt{2} \, \tr\big( [\phi^\dag, \phi ] \, \bar{\zeta} \, \bar{\xi} \big)
	\nonumber \\ &&
	- \sqrt{2} \, i \, \tr\big( \phi \, D_\mu \bar{\chi} \,
	\bar{\sigma}^\mu \, \zeta \big)
	- \sqrt{2} \, i \, \tr\big( \phi^\dag \, \bar{\zeta} \,
	\bar{\sigma}^\mu \, D_\mu \chi \big)
	- \sqrt{2} \, \tr\big( F^\dag \, \zeta \, \chi \big)
	- \sqrt{2} \, \tr\big( F \, \bar{\zeta} \, \bar{\chi} \big)
	\nonumber \\ &&
	+ \frac{1}{\sqrt{2}} \, f \, \tr\big( F^\dag \phi \big)
	+ \frac{1}{\sqrt{2}} \, f^* \, \tr\big( \phi^\dag F \big)
	+ \frac{1}{\sqrt{2}} \, f \, \tr\big( \xi^2 \big)
	+ \frac{1}{\sqrt{2}} \, f^* \, \tr\big( \bar{\xi}^2 \big),
	\label{eq:L:components:matter}
\end{eqnarray}
where the field strength tensor is $F_{\mu\nu} = \partial_\mu A_\nu - \partial_\nu A_\mu - i [ A_\mu, A_\nu]$ and its dual $\tilde{F}^{\mu\nu} = \frac{1}{2} \epsilon^{\mu\nu\rho\sigma} F_{\rho\sigma}$.
The first three lines coincide with the Lagrangian of the ordinary $\N = 2$ pure gauge theory with vanishing $\theta$-term; notice however that we have used integration by parts to remove all derivatives acting on the coupling field $g$, and therefore it not optional to write for instance the kinetic term for $\phi$ with second derivatives instead of a term like $\tr( D_\mu \phi^\dag \, D^\mu \phi)$.
When $\theta$ is a constant, the fourth line is a total derivative and vanishes in the action. All the remaining terms, linear in $\zeta$ or $f$, are needed to enforce invariance of the action under the supersymmetry transformation that takes $g$ and $\theta$ into their superpartners.
There is only one supersymmetry in the Lagrangian \eqref{eq:L:components:matter}, even though the part proportional to $g$ could seem to be invariant under a full extended $\N = 2$ supersymmetry. This is because the superspace Lagrangian~\eqref{eq:L:superspace} we begin with does not preserve this extended supersymmetry. It would be relatively easy to restore it in eq.~\eqref{eq:L:components:matter} by adding $\N = 2$ superpartners for the coupling fields: $\theta$ would become part of a triplet of the $SU(2)$ R-symmetry, while $g$ would remain a singlet, explaining the fact that it couples to a manifestly R-symmetric term in the action. What would be more complicated to obtain in an explicitly extended supersymmetric setup is the logarithmic dependence on the cutoff of the action \eqref{eq:L:superspace:cutoff}. Since we do not make any use of extended supersymmetry in this work, it is sufficient for all our purposes to consider the Lagrangian written above. Moreover, we will later set all coupling fields to zero with the exception of the R-singlet $g$, implying that our results will actually be valid in a complete $\N = 2$ supersymmetric setup.

For the vacuum part of the action, we do not bother writing down the terms proportional to the fermionic partner $\zeta$ of the coupling $g$, as there are many of them and they do not play any role in our analysis. We have therefore
\begin{eqnarray}
	\L_\text{vacuum} & = & 
	\frac{1}{2} \, \square g \, \square g
	+ \frac{1}{2} \, \square \theta \, \square \theta
	+ \partial_\mu f^* \, \partial^\mu f
	\nonumber \\
	&& + \frac{2}{3} \left( \frac{1}{2} \, \partial_\mu g \, \partial^\mu g
	+ \frac{1}{2} \, \partial_\mu \theta \, \partial^\mu \theta
	+ f^* f \right)^2
	\nonumber \\
	&&
	- \frac{2}{3} \, \partial_\mu g \, \partial^\mu g \,
	\partial_\nu \theta \, \partial^\nu \theta
	+ \frac{2}{3} \, \partial_\mu g \, \partial_\nu g \,
	\partial^\mu \theta \, \partial^\nu \theta
	\nonumber \\
	&& +~\text{terms in}~\zeta, \bar{\zeta}
	\label{eq:L:components:vacuum}
\end{eqnarray}
Eq.~\eqref{eq:L} is nothing but the Lagrangian of eqs.~\eqref{eq:L:components:matter} and \eqref{eq:L:components:vacuum} when the auxiliary fields are integrated out and when all the coupling fields are set to zero, with the exception of $g$. Setting the fermionic component of the coupling superfield to zero makes sense in a search for solitons, as it cannot acquire a vacuum expectation value and its equations of motion is trivially satisfied by vanishing fields. For the bosonic components $\theta$ and $f$, we present below the reasoning behind that choice.

\subsection{The equations of motion in components}

The equations of motion that correspond to the components of the superfield equations~(\ref{eq:superfieldeq:Gbar}--\ref{eq:superfieldeq:V}) can be derived from the Lagrangian in components, setting all fermionic field to zero for simplicity.
$D$ and $F$ are auxiliary fields satisfying
\begin{equation}
	\frac{\delta S}{\delta D}
	= b \, g \, \left( 2 \, D - [ \phi^\dag, \phi ] \right),
	\hspace{1.5cm}
	\frac{\delta S}{\delta F^\dag}
	= b \, \left( g \, F
	+ \frac{1}{\sqrt{2}} \, f \, \phi \right),
	\label{eq:eom:DF}
\end{equation}
Since the dynamics of these fields is trivially fixed, they can be directly replaced by the solution to their equations of motion in the action
The gauge and matter fields obey the usual equations, augmented with terms involving derivatives of the couplings $g$ and $\theta$, as well as $|f|^2$:
\begin{equation}
	\frac{\delta S}{\delta A^\mu}
	= - b \, \left( g \, D^\nu F_{\mu\nu}
	+ \frac{i}{2} \, g \, \phi^\dag \, \overleftrightarrow{D_\mu} \phi
	+ \frac{i}{2} \, g \, \phi \, \overleftrightarrow{D_\mu} \phi^\dag
	+ \partial^\nu g \, F_{\mu\nu}
	+ \partial^\nu \theta \, \tilde{F}_{\mu\nu}
	- \frac{1}{2} \, \partial_\mu \theta \, [ \phi^\dag, \phi ] \right),
	\label{eq:eom:A:complete}
\end{equation}
\begin{equation}
	\frac{\delta S}{\delta \phi^\dag}
	= -\frac{b}{2} \bigg( g \, D_\mu D^\mu \phi
	+ \frac{1}{2} \, g \, \big[ \phi, [ \phi^\dag, \phi ] \big]
	+ \partial_\mu \left( g + i \theta \right) \, D^\mu \phi
	+ \frac{1}{2} \, \left[ \square \left( g + i \theta \right)
	+ \frac{|f|^2}{g} \right] \, \phi \bigg).
	\label{eq:eom:phi:complete}
\end{equation}
The equations of motion for the couplings $g$ and $\theta$ are 
\begin{eqnarray}
	\frac{\delta S}{\delta g}
	& = & - \frac{b}{2} \, \tr\big( F_{\mu\nu} F^{\mu\nu} \big)
	- \frac{b}{2} \, \tr\big( \phi^\dag D_\mu D^\mu \phi \big)
	- \frac{b}{2} \, \tr\big( \phi \, D_\mu D^\mu \phi^\dag \big)
	- \frac{b}{4} \, \tr\big( [ \phi^\dag, \phi ]^2 \big)
	\nonumber \\
	&& + \frac{b}{2} \, \frac{f^\dag f}{g^2} \tr\big( \phi^\dag \phi \big)
	+ c \, \square^2 g
	- \frac{4c}{3} \, \square g \,
	\left( \frac{1}{2} \, \partial_\mu g \, \partial^\mu g
	- \frac{1}{2} \, \partial_\mu \theta \, \partial^\mu \theta
	+ |f|^2 \right)
	\nonumber \\
	&& - \frac{4c}{3} \, \partial_\mu \partial_\nu g \,
	\left( \partial^\mu g \, \partial^\nu g
	+ \partial^\mu \theta \, \partial^\nu \theta \right)
	- \frac{4c}{3} \, \partial_\mu g \,
	\left( \partial^\mu \theta \, \square \theta
	+ f^* \partial^\mu f + f \, \partial^\mu f^* \right) ,
	\label{eq:eom:g:complete}
	\\
	\frac{\delta S}{\delta \theta}
	& = & - \frac{b}{2} \,
	\tr\big( F_{\mu\nu} \tilde{F}^{\mu\nu} \big)
	+ \frac{i \, b}{2} \, \partial^\mu 
	\tr\big( \phi^\dag \overleftrightarrow{D_\mu} \phi \big)
	\nonumber \\
	&& + c \, \square^2 \theta
	- \frac{4c}{3} \, \square \theta \,
	\left( -\frac{1}{2} \, \partial_\mu g \, \partial^\mu g
	+ \frac{1}{2} \, \partial_\mu \theta \, \partial^\mu \theta
	+ |f|^2 \right)
	\nonumber \\
	&& - \frac{4c}{3} \, \partial_\mu \partial_\nu \theta \,
	\left( \partial^\mu g \, \partial^\nu g
	+ \partial^\mu \theta \, \partial^\nu \theta \right)
	- \frac{4c}{3} \, \partial_\mu \theta \, \left( \partial^\mu g \, \square g
	+ f^* \partial^\mu f + f \, \partial^\mu f^* \right).
	\label{eq:eom:theta}
\end{eqnarray}
Finally, the equation of motion for the complex field $f$ reads
\begin{equation}
	\frac{\delta S}{\delta f^*}
	= -\frac{b}{2} \, \frac{f}{g} \, \tr\big( \phi^\dag \phi \big)
	- c \, \square f
	+ \frac{4c}{3} \, f \, 
	\left( \frac{1}{2} \, \partial_\mu g \, \partial^\mu g
	+ \frac{1}{2} \, \partial_\mu \theta \, \partial^\mu \theta
	+ |f|^2 \right).
	\label{eq:eom:f}
\end{equation}

\subsection{The supercurrent and energy-momentum tensor in components}

The conserved $R$-current associated with the supersymmetry that is manifest in our $\N = 1$ superspace formalism is given by the lowest component of the supercurrent $J^\mu = \left( \bar{\sigma}^\mu \right)^{\alphadot\alpha} J_{\alpha\alphadot}$, namely
\begin{eqnarray}
	j^\mu & = & b \, \left[
	\frac{i \, g}{6} \,
	\tr\big( \phi \, \overleftrightarrow{D^\mu} \phi^\dag \big)
	- \frac{g}{2} \, \bar{\xi} \, \sigma^\mu \, \xi
	+ \frac{g}{6} \, \bar{\chi} \, \sigma^\mu \, \chi
	- \frac{1}{6} \, \partial^\mu \theta \, \tr\big( \phi^\dag \phi \big)
	\right]
	\nonumber \\
	&& - c \, \left[ \frac{1}{3} \, \partial_\nu
	\left( \partial^\mu g \, \partial^\nu \theta
	- \partial^\nu g \, \partial^\mu \theta \right)
	+ \frac{i}{2} \, f^* \overleftrightarrow{\partial^\mu} f \right]
	+ \Big[ \text{terms in}~\zeta, \bar{\zeta} \Big]
\end{eqnarray}
The energy momentum tensor can similarly be extracted from the $\theta\bar{\theta}$ component of the supercurrent, giving
\begin{eqnarray}
	T^{\mu\nu} & = & b \bigg[ \frac{1}{2} \, \eta^{\mu\nu} \,
	g \, \tr\big( F_{\rho\sigma} F^{\rho\sigma} \big)
	- 2 \, g \, \tr\big( F^\mu_{~\rho} F^{\nu\rho} \big)
	- \frac{1}{12} \, \eta^{\mu\nu} \, g \,
	\tr\big( [ \phi^\dag, \phi ]^2 \big)
	\nonumber \\
	&& \quad
	- \frac{1}{3} \, \eta^{\mu\nu} \, g \, 
	\tr\big( D_\rho \phi^\dag \, D^\rho \phi \big)
	+ \frac{2}{3} \, g \,
	\tr\big( D^\mu \phi^\dag \, D^\nu \phi \big)
	+ \frac{2}{3} \, g \,
	\tr\big( D^\nu \phi^\dag \, D^\mu \phi \big)
	\nonumber \\
	&& \quad
	- \frac{1}{3} \, g \,
	\tr\big( \phi^\dag \, D^\mu D^\nu \phi \big)
	- \frac{1}{3} \, g \,
	\tr\big( \phi \, D^\mu D^\nu \phi^\dag \big)
	- \frac{1}{3} \, \partial^\mu \partial^\nu g \,
	\tr\big( \phi^\dag \phi \big)
	\nonumber \\
	&& \quad
	- \frac{1}{6} \, \eta^{\mu\nu} \, \partial_\rho g \,
	\tr\big( \phi^\dag D^\rho \phi \big)
	+ \frac{1}{6} \, \partial^\mu g \,
	\tr\big( \phi^\dag D^\nu \phi \big)
	+ \frac{1}{6} \, \partial^\nu g \,
	\tr\big( \phi^\dag D^\mu \phi \big)
	\nonumber \\
	&& \quad
	- \frac{1}{6} \, \eta^{\mu\nu} \, \partial_\rho g \,
	\tr\big( \phi \, D^\rho \phi^\dag \big)
	+ \frac{1}{6} \, \partial^\mu g \,
	\tr\big( \phi \, D^\nu \phi^\dag \big)
	+ \frac{1}{6} \, \partial^\nu g \,
	\tr\big( \phi \, D^\mu \phi^\dag \big)
	\nonumber \\
	&& \quad
	+ \frac{i}{6} \, \eta^{\mu\nu} \, \partial_\rho \theta \,
	\tr\big( \phi^\dag \, \overleftrightarrow{D^\rho} \phi \big)
	- \frac{i}{2} \, \partial^\mu \theta \,
	\tr\big( \phi^\dag \, \overleftrightarrow{D^\nu} \phi \big)
	- \frac{i}{2} \, \partial^\nu \theta \,
	\tr\big( \phi^\dag \, \overleftrightarrow{D^\mu} \phi \big)
	\nonumber \\
	&& \quad
	+ \frac{1}{6} \, \eta^{\mu\nu} \, \frac{|f|^2}{g} \,
	\tr\big( \phi^\dag \phi \big) \bigg]
	\nonumber \\
	&& + c \bigg[
	- \frac{1}{2} \, \eta^{\mu\nu} \, \square g \, \square g
	+ 2 \, \partial^\mu \partial^\nu g \, \square g
	+ \frac{1}{3} \, \eta^{\mu\nu} \, \partial_\rho \partial_\sigma g \,
	\partial^\rho \partial^\sigma g
	- \frac{4}{3} \, \partial^\mu \partial_\rho g \,
	\partial^\nu \partial^\rho g
	\nonumber \\
	&& \qquad
	+ \frac{1}{3} \, \eta^{\mu\nu} \, \partial_\rho g \,
	\partial^\rho \square g
	- \partial^\mu g \, \partial^\nu \square g
	- \partial^\nu g \, \partial^\mu \square g
	+ \frac{2}{3} \, \partial_\rho g \,
	\partial^\mu \partial^\nu \partial^\rho g
	\nonumber \\
	&& \qquad
	- \frac{1}{2} \, \eta^{\mu\nu} \, \square \theta \, \square \theta
	+ 2 \, \partial^\mu \partial^\nu \theta \, \square \theta
	+ \frac{1}{3} \, \eta^{\mu\nu} \, \partial_\rho \partial_\sigma \theta \,
	\partial^\rho \partial^\sigma \theta
	- \frac{4}{3} \, \partial^\mu \partial_\rho \theta \,
	\partial^\nu \partial^\rho \theta
	\nonumber \\
	&& \qquad
	+ \frac{1}{3} \, \eta^{\mu\nu} \, \partial_\rho \theta \,
	\partial^\rho \square \theta
	- \partial^\mu \theta \, \partial^\nu \square \theta
	- \partial^\nu \theta \, \partial^\mu \square \theta
	+ \frac{2}{3} \, \partial_\rho \theta \,
	\partial^\mu \partial^\nu \partial^\rho \theta
	\nonumber \\
	&& \qquad
	- \frac{1}{3} \, \eta^{\mu\nu} \, \partial_\rho f^* \, \partial^\rho f
	+ \frac{2}{3} \, \partial^\mu f^* \, \partial^\nu f
	+ \frac{2}{3} \, \partial^\nu f^* \, \partial^\mu f
	\nonumber \\
	&& \qquad
	+ \frac{1}{3} \, \eta^{\mu\nu} \, f^* \, \square f
	- \frac{1}{3} \, f^* \, \partial^\mu \partial^\nu f
	+ \frac{1}{3} \, \eta^{\mu\nu} \, f \, \square f^*
	- \frac{1}{3} \, f \, \partial^\mu \partial^\nu f^*
	\nonumber \\
	&& \qquad
	- \frac{1}{6} \, \eta^{\mu\nu} \,
	\left( \partial_\rho g \, \partial^\rho g \right)^2
	+ \frac{2}{3} \, \partial^\mu g \, \partial^\nu g \,
	\partial_\rho g \, \partial^\rho g
	- \frac{1}{6} \, \eta^{\mu\nu} \,
	\left( \partial_\rho \theta \, \partial^\rho \theta \right)^2
	+ \frac{2}{3} \, \partial^\mu \theta \, \partial^\nu \theta \,
	\partial_\rho \theta \, \partial^\rho \theta
	\nonumber \\
	&& \qquad
	+ \frac{1}{3} \, \eta^{\mu\nu} \, \partial_\rho g \, \partial^\rho g \,
	\partial_\sigma \theta \, \partial^\sigma \theta
	- \frac{2}{3} \, \partial^\mu g \, \partial^\nu g \,
	\partial_\rho \theta \, \partial^\rho \theta
	- \frac{2}{3} \, \partial_\rho g \, \partial^\rho g \,
	\partial^\mu \theta \, \partial^\nu \theta
	\nonumber \\
	&& \qquad
	- \frac{2}{3} \, \eta^{\mu\nu} \, \partial_\rho g \, \partial_\sigma g \,
	\partial^\rho \theta \, \partial^\sigma \theta
	+ \frac{4}{3} \, \partial^\mu g \, \partial_\rho g \,
	\partial^\nu \theta \, \partial^\rho \theta
	+ \frac{4}{3} \, \partial^\nu g \, \partial_\rho g \,
	\partial^\mu \theta \, \partial^\rho \theta
	\nonumber \\
	&& \qquad
	- \frac{2}{3} \, \eta^{\mu\nu} \, \partial_\rho g \, \partial^\rho g \,
	|f|^2
	+ \frac{4}{3} \, \partial^\mu g \, \partial^\nu g \, |f|^2
	- \frac{2}{3} \, \eta^{\mu\nu} \, \partial_\rho \theta \,
	\partial^\rho \theta \, |f|^2
	+ \frac{4}{3} \, \partial^\mu \theta \, \partial^\nu \theta \, |f|^2
	\nonumber \\
	&& \qquad
	- \frac{2}{3} \, \eta^{\mu\nu} \, \left( |f|^2 \right)^2 
	\bigg] + \Big[ \text{terms containing fermionic fields} \Big].
	\label{eq:T:complete}
\end{eqnarray}
Both the current $j^\mu$ and the energy-momentum tensor $T^{\mu\nu}$ are conserved upon the equations of motion, and the energy-momentum tensor is moreover traceless.

If we restrict the our study to static configurations (independent of time) and choose to work in the temporal gauge $A_0 = 0$, the energy density is given by
\begin{eqnarray}
	T^{00} & = & b \, \bigg[ \frac{1}{2} \,
	g \, \tr\big( F_{ij} F_{ij} \big)
	- \frac{1}{12} \, g \, \tr\big( [\phi^\dag, \phi]^2 \big)
	+ \frac{1}{3} \, g \, 
	\tr\big( D_i \phi^\dag \, D_i \phi \big)
	\nonumber \\
	&& \quad
	+ \frac{1}{6} \, \partial_i g \, \partial_i 
	\tr\big( \phi^\dag \phi \big)
	- \frac{i}{6} \, \partial_i \theta \,
	\tr\big( \phi^\dag \, \overleftrightarrow{D_i} \phi \big)
	+ \frac{1}{6} \, \frac{|f|^2}{g} \, \tr\big( \phi^\dag \phi \big)
	\bigg]
	\nonumber \\
	&& + c \, \bigg[
	- \frac{1}{2} \, \partial^2 g \, \partial^2 g
	+ \frac{1}{3} \, \partial_i \partial_j g \, \partial_i \partial_j g
	+ \frac{1}{3} \, \partial_i g \, \partial_i \partial^2 g
	\nonumber \\
	&& \qquad
	- \frac{1}{2} \, \partial^2 \theta \, \partial^2 \theta
	+ \frac{1}{3} \, \partial_i \partial_j \theta \,
	\partial_i \partial_j \theta
	+ \frac{1}{3} \, \partial_i \theta \,
	\partial_i \partial^2 \theta
	\nonumber \\
	&& \qquad
	- \frac{1}{6} \, \left( \partial_i g \, \partial_i g \right)^2
	- \frac{1}{6} \,
	\left( \partial_i \theta \, \partial_i \theta \right)^2
	+ \frac{1}{3} \, \partial_i g \, \partial_i g \,
	\partial_j \theta \, \partial_j \theta
	- \frac{2}{3} \, \partial_i g \, \partial_j g \,
	\partial_i \theta \, \partial_j \theta
	\nonumber \\
	&& \qquad
	+ \frac{1}{3} \, \partial_i f^* \, \partial_i f
	- \frac{1}{3} \, f^* \, \partial^2 f
	- \frac{1}{3} \, f \, \partial^2 f^*
	\nonumber \\
	&& \qquad
	+ \frac{2}{3} \, \left( \partial_i g \, \partial_i g
	+ \partial_i \theta \, \partial_i \theta \right) \, |f|^2
	- \frac{2}{3} \, |f|^4 
	\bigg].
	\label{eq:T00:complete}
\end{eqnarray}
Making use of the equations of motion, all the dependence on the gauge and matter fields can be turned into a boundary term:
\begin{eqnarray}
	T^{00} & = & c \, \bigg[ \frac{1}{2} \, \left( \partial^2 g \right)^2
	+ \frac{1}{2} \, \left( \partial^2 \theta \right)^2
	+ \frac{1}{2} \left( \partial_i g \, \partial_i g
	- \partial_i \theta \, \partial_i \theta \right)^2
	+ 2 \, \left( \partial_i g \, \partial_i \theta \right)^2
	\nonumber \\
	&& \quad
	-  \partial_i f^* \, \partial_i f
	- 2 \, |f|^2 \, \left( \partial_i g \, \partial_i g
	+ \partial_i \theta \, \partial_i \theta \right)
	+ 2 \, |f|^4 \bigg]
	+ \partial_i \mathcal{B}_i,
	\label{eq:T00:positive}
\end{eqnarray}
where we have defined
\begin{eqnarray}
	\mathcal{B}_i & = & b \left[
	\frac{1}{6} \,g \, \partial_i \tr\big( \phi^\dag \phi \big)
	- \frac{1}{3} \, \partial_i g \, \tr\big( \phi^\dag \phi \big)
	- \frac{i}{2} \, \theta \,
	\tr\big( \phi^\dag \overleftrightarrow{D_i} \phi \big) \right]
	\nonumber \\
	&& + c \, \bigg[ g \, \partial_i \partial^2 g
	- \partial_i g \, \partial^2 g
	+ \frac{1}{3} \, \partial_j g \, \partial_i \partial_j g
	+ \theta \, \partial_i \partial^2 \theta
	- \partial_i \theta \, \partial^2 \theta
	+ \frac{1}{3} \, \partial_j \theta \, \partial_i \partial_j \theta
	\nonumber \\
	&& \qquad
	- \frac{2}{3} \, \left( g \, \partial_i g - \theta \, \partial_i \theta \right)
	\left( \partial_j g \, \partial_j g - \partial_j \theta \, \partial_j \theta \right)
	- \frac{4}{3} \, g \, \partial_j g \,
	\partial_i \theta \, \partial_j \theta
	- \frac{4}{3} \, \partial_i g \,
	\partial_j g \, \theta \, \partial_j \theta
	\nonumber \\
	&& \qquad
	+ \, \frac{2}{3} \, \partial_i |f|^2
	+ \frac{4}{3} \, \left( g \, \partial_i g
	+ \theta \, \partial_i \theta \right) |f|^2 \bigg].
\end{eqnarray}
This boundary terms does not contribute to the energy of a static configuration, provided that the fields and couplings approach a constant value at spatial infinity.
In this case, it can be seen from eq.~\eqref{eq:T00:positive} that the energy density is a sum of positive terms, with the notable exception of terms that contain the field $f$.

\begin{figure}
	\centering
	\includegraphics[scale=0.5]{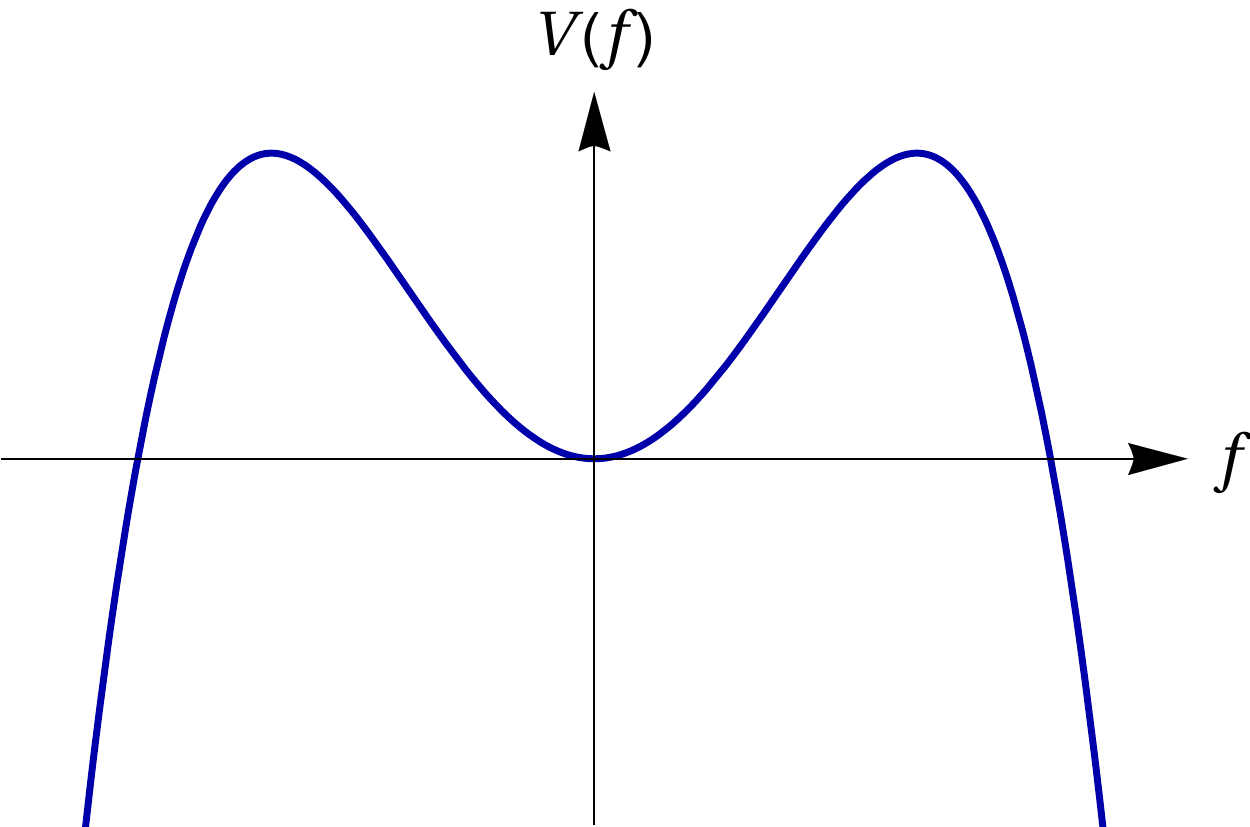}
	\caption{The inverted Higgs potential for $f$, with its local minimum at $f=0$.}
	\label{fig:fpotential}
\end{figure}%
To understand the role of $f$, it is best to look back at the action in components as given by eqs.~\eqref{eq:L:components:matter} and \eqref{eq:L:components:vacuum}: the action for $f$ is that of an ordinary scalar field with a $|f|^4$ self-interaction. It turns out however that the sign of this self-interaction term is ``wrong'', in the sense that it leads to a potential that is unbounded below, hence the presence of terms proportional to $f$ that are negative in the energy density~\eqref{eq:T00:positive}. Still, the quantity that plays the role of the mass term for $f$ is non-negative for static configurations, as seen in eq.~\eqref{eq:T00:complete}, and thus the potential has generically a local minimum close to the trivial solution $f = 0$, as illustrated in Fig.~\ref{fig:fpotential}. 
In this work, we choose therefore to limit our analysis to configurations in which $f$ vanishes identically.
In this case, the energy is bounded below under the relatively mild assumption that the couplings tend to a constant at spatial infinity and that the matter field has a constant vacuum expectation value.

Finally, we also choose deliberately to limit our analysis to the solution with $\theta = 0$. As can be seen from eq.~\eqref{eq:eom:theta}, this is not a solution to its equation of motion, but rather an external assumptions, which is actually equivalent to the vanishing of a topological current.
Let us remark once again that our approach explicitly breaks $\N = 2$ supersymmetry, and that the dynamics of the fields $\theta$ and $f$ could actually be slightly different in a completely $\N = 2$ supersymmetric treatment of the problem. Such a treatment is postponed for future work.


\bibliography{Bibliography}{}
\bibliographystyle{JHEP}

\end{document}